\documentclass[12pt,twoside,a4paper]{article}
\usepackage[utf8]{inputenc}
\usepackage{graphicx,color}
\usepackage{times}
\usepackage{xspace}
\usepackage{booktabs}
\usepackage[english]{babel}
\usepackage{amssymb}
\usepackage{cite}
\usepackage{xspace}
\usepackage{a4wide}
\usepackage{amsmath,multicol}
\usepackage[toc]{multitoc}
\usepackage{subcaption}

\usepackage{afterpage}
\usepackage{multirow}
\usepackage{url}

\usepackage[colorlinks=true, pdftex, citecolor=blue, urlcolor=blue, linkcolor=blue]{hyperref}

\newcommand{\mrm}[1]{\ensuremath{\mathrm{#1}}\xspace}

\newcommand{\Rt}{\ensuremath{R_{\rm T}}\xspace}

\newcommand{\Nch}{\ensuremath{N_{\rm Inc.}}\xspace}

\newcommand{\kshort}{\ensuremath{K_{\rm s}^0}\xspace}

\newcommand{\hw}{\textsc{Herwig}\xspace}
\newcommand{\pyNonum}{\textsc{Pythia}\xspace}
\newcommand{\py}{\textsc{Pythia 8}\xspace}
\newcommand{\sh}{\textsc{Sherpa}\xspace}
\newcommand{\dipsy}{\textsc{Dipsy}\xspace}
\newcommand{\dipsyrope}{\textsc{Dipsy Rope}\xspace}
\newcommand{\dipsynoswing}{\textsc{Dipsy NoSwing}\xspace}
\newcommand{\epos}{\textsc{epos}\xspace}

\newcommand{\MeV}{\ifmmode {\mathrm{\ Me\kern -0.1em V}}\else
                   \textrm{Me\kern -0.1em V}\fi}%
\newcommand{\GeV}{\ifmmode {\mathrm{\ Ge\kern -0.1em V}}\else
                   \textrm{Ge\kern -0.1em V}\fi}%
\newcommand{\TeV}{\ifmmode {\mathrm{\ Te\kern -0.1em V}}\else
                   \textrm{Te\kern -0.1em V}\fi}%

\newcommand{\pt}{\ensuremath{p_{\perp}}\xspace}

\newcommand{\eqRef}[1]{eq.~(\ref{#1})\xspace}

\newcommand{\secRef}[1]{section~\ref{#1}\xspace}

\newcommand{\tabRef}[1]{tab.~\ref{#1}\xspace}

\newcommand{\figRef}[1]{fig.~\ref{#1}\xspace}

\renewcommand{\and}{, }

\begin{document}
\vspace*{-1.75cm}\begin{minipage}{\textwidth}
\flushright\small
COEPP-MN-16-6\\
MCNET-16-06
\end{minipage}
\vskip1.25cm
{\Large\bf
\begin{center}
Probing Collective Effects in Hadronisation with the Extremes of the Underlying Event
\end{center}}
\vskip5mm
{\begin{center}
{\large
 T.~Martin$^1$, P.~Skands$^{2}$, S.~Farrington$^1$
}\end{center}
$^1$:~\parbox[t]{0.985\textwidth}{Department of Physics, University of Warwick, CV47AL, UK}\\
$^2$:~\parbox[t]{0.985\textwidth}{School of Physics and Astronomy, Monash University, Clayton VIC-3800, Australia}\\[3mm]
\vskip5mm
\begin{center}
\parbox{0.88\textwidth}{
\begin{center}
\textbf{Abstract}
\end{center}\small
We define a new set of observables to probe the structure of the underlying event in hadron collisions. We use the conventional definition of the ``transverse region'' in jet events and, for a fixed window in jet $p_\perp$, propose to measure several discriminating quantities as a function of the level of activity in the transverse region. The measurement of these observables in LHC data would reveal whether, e.g., the properties of ``low-UE'' events are compatible with equivalent measurements in $e^+e^-$ collisions (jet universality), and whether the scaling behaviour towards ``high-UE'' events exhibits properties of non-trivial soft-QCD dynamics, such as colour re-connections or other collective phenomena. We illustrate at $\sqrt{s} = 13 \TeV$ that significant discriminatory power is obtained in comparisons between MC models with varying treatments of collective effects, including \py, \epos, and \dipsy.
}
\end{center}\vspace*{2mm}}

\section{Introduction \label{sec:intro}}
The ``jet pedestal effect'', now called the underlying event (UE), was first observed by the UA1 experiment at CERN's $\mathrm{Sp\bar{p}S}$ collider, 
in an early study of jet events~\cite{Arnison:1983gw}. This study concluded that, outside the core of a jet, ``a constant $E_{\rm T}$ plateau is observed, whose height is independent of the jet $E_{\rm T}$''~\cite{Arnison:1983gw}. The plateau was substantially higher than that observed in minimum-bias events\footnote{For a definition of the term minimum-bias, see e.g.~\cite[Chp.~7]{Buckley:2011ms}.} at the same energy~\cite{Arnison:1982ds,Arnison:1981ks,Arnison:1982rm,Albajar:1988tt}. Its characteristics have since been studied extensively, in $pp$ and $p\bar{p}$ collisions over two orders of magnitude in Centre of Mass (CM) energy, from RHIC~\cite{Caines:2009iy,Caines:2011zza} through Tevatron~\cite{Affolder:2001xt,Acosta:2004wqa,Aaltonen:2010rm,Aaltonen:2015aoa} to LHC energies~\cite{Khachatryan:2010pv,Aad:2010fh,Aad:2011qe,Chatrchyan:2011id,ALICE:2011ac,Aad:2012mfa,Chatrchyan:2012tt,Chatrchyan:2012tb,Chatrchyan:2013gfi,Chatrchyan:2013ala,Aad:2014hia,Aad:2014jgf,Khachatryan:2015jza}, the latter most recently extending to measurements at 13 TeV~\cite{ATL-PHYS-PUB-2015-019}.

Theoretically, the UE is often roughly defined as the collection of particles produced in a single hadronic interaction which do \emph{not} originate from the primary (``hard'') parton-parton scattering, nor relate directly to it in the form of hadronised parton showers. However, since the hard-interaction initiators are coloured (quarks or gluons), confinement implies that it is in general not possible to define uniquely which hadrons were produced by the hard interaction and which belong to the UE. 

Instead, physical observables that measure phase-space regions ``outside the core of jets'' are used to provide operational definitions of the UE. Two main definitions are in current use: the most common one is a purely geometrical definition of the UE as an angular region transverse to a leading charged particle or jet~\cite{Affolder:2001xt}, typically spanning $90^\circ \pm 30^\circ$; an alternative definition of the UE is what remains when hard jets have been removed, an early example of this being the ``Swiss Cheese'' method introduced by CDF~\cite{Acosta:2004wqa} now superseded by so-called ``jet median/area'' techniques proposed in~\cite{Cacciari:2009dp} and used e.g.\ by CMS~\cite{Heinrich:2011bla,Chatrchyan:2012tt}.

In addition to the average UE properties, several studies at the LHC have now also measured observables sensitive to the per-event \emph{fluctuations}. These include the ATLAS \cite{Aad:2010fh} study of the widths of the charged-particle density and scalar-\pt sum distributions, and the CMS \cite{Chatrchyan:2013ala} study of jet multiplicity in the UE. Such studies are crucial to pin down the underlying physics mechanisms, revealing e.g.\ potentially interesting deviations from Poissonian fluctuations, and to what extent (soft) jets are required to build up the UE activity. In this paper, we propose to go a step further, and study the UE properties as a function of its in-situ level; this will allow a detailed exploration of the evolution of quantities such as strangeness fractions, particle spectra, and (soft) jet rates, from low to high underlying activity. 
Though we focus mainly on identified-particle rates, we emphasise that the method is general and can be applied to any UE property as a function of any measure of UE activity. The explicit studies we propose are inspired by --- and are intended to be complementary to --- minimum-bias studies as a function of multiplicity. Of special interest is whether events with very low UE levels exhibit particle yields and spectra more consistent with LEP fragmentation models than those of their higher-UE counterparts, and whether high-UE events exhibit any clear signs of flow or other collective effects. 
A complementary recent interesting proposal is that of~\cite{Bierlich:2015rha} which suggests to measure similar observables using forward multiplicity as the indicator of event activity.

On the theory side, the fragmentation process in the simple reference
case of hadronic $Z$ decays is believed to be relatively well
understood, and shower+hadronisation models such as those implemented
in the general-purpose Monte Carlo models \hw~\cite{Bahr:2008pv},
\pyNonum~\cite{Sjostrand:2014zea}, and \sh~\cite{Gleisberg:2008ta},
are capable of describing the majority of $e^+e^-\to\mrm{hadrons}$
data well (see e.g.\ the plots available on
\href{http://mcplots.cern.ch/?query=plots,ee}{mcplots}~\cite{Karneyeu:2013aha}). However,
moving to $pp$ collisions, several additional complications arise,
including initial-state radiation, beam remnants, diffraction, and
multiple parton interactions (MPI). The corresponding physics models
are therefore necessarily also more complicated. Starting from Rick Field's observation with his ``Tune A''~\cite{Field:2005sa} of the
\pyNonum~6 MPI model~\cite{Sjostrand:1987su,Sjostrand:2006za} to
Tevatron data~\cite{Acosta:2004wqa} that a much higher degree of
colour correlation than anticipated (between partons from different MPI systems) was required to fit the $p_\perp$ spectra of particles in the UE at CDF, there have appeared
progressively more indications that the physics of hadron collisions
is more complicated than was previously thought. The clues include the
increase of the average $p_\perp$ with event
multiplicity~\cite{McLerran:1986nc,Sjostrand:1987su,Albajar:1989an,Acosta:2001rm,Aaltonen:2009ne,Aad:2010ac,Khachatryan:2010nk},
the large yields of
hyperons~\cite{Adams:2006yu,Aaltonen:2011wz,Aad:2011hd,Khachatryan:2011tm,Aaij:2012ut,Abelev:2012jp,Chatrchyan:2013qsa,Abelev:2014qqa},
and the overall relatively steep increase of average $p_\perp$ with
hadron
mass~\cite{Abelev:2006cs,Aad:2011hd,Aamodt:2011zj,Khachatryan:2011tm,Abelev:2012hy,Chatrchyan:2013qsa,Abelev:2014qqa,Adam:2015qaa}
observed in $pp$ collisions. There are also more subtle indications, such as the unexpected ridge-like structure in two-particle correlations at low $\Delta\phi$ in high-multiplicity
minimum-bias $pp$ collisions observed by CMS and ATLAS~\cite{Khachatryan:2010gv,Aad:2015gqa}. To shed
further light on these effects, many studies of minimum-bias events
now focus on the scaling with charged-particle multiplicity (see e.g.\
\cite{Acosta:2005pk,Aad:2010ac,Khachatryan:2010nk,Abelev:2013sqa,Bierlich:2015rha,Ortiz:2016mra}),
with a special emphasis on the highest accessible multiplicities. Here we extend this type of study to the UE context.   

At central rapidities, particle production in the UE is predominantly
driven by the hadronisation of soft/semi-hard secondary scatters
within the colliding hadrons. The presence of the requirement for high-$p_\perp$ activity, such as that provided by a (hard) experimental trigger, biases the UE
towards larger activities than that of more inclusive (e.g.\ minimum-bias) event samples. Within the context of MPI models (see e.g.~\cite[Chp.~7]{Buckley:2011ms}), this can
be understood as a consequence of a bias towards smaller impact
parameters, with the onset of the aforementioned UE plateau signaling that 
the average impact parameter has become significantly smaller
than the radius of the proton $\left<b\right> \ll r_p$, such that the
two matter distributions are already completely overlapping from that
point onward (see e.g.~\cite[sect.~VI.D]{Sjostrand:1987su}). There are, however, still large fluctuations expected,
some geometric in origin, others dynamic. Examples of geometric
fluctuations include, for instance, the rare case of a large-impact
parameter interaction producing a hard jet presumably 
accompanied by very little UE, or the case of two protons happening to be
compressed transversely producing a larger-than-ordinary matter
overlap and hence a very large UE. Even for fixed matter overlap, the
number and hardness of MPI and shower branchings will exhibit random 
(quantum) fluctuations, and especially the tail towards high activities may
be sensitive to new QCD dynamics. Currently, we are not aware of 
any single phenomenological model that spans all the
possibilities. Nevertheless, on general grounds, we  expect baryon and strangeness fractions to be good tracers of the underlying mechanisms (in addition to hard MPI rates):
\begin{itemize}
\item The strange quark, having
a mass very near 
$\Lambda_\mrm{QCD}$,  can act as a sensitive probe of any
changes to the energy density available at the time of
hadronisation. Simply put, the more strange quarks we see, relative to $u$ and $d$ quarks, the higher the effective string tension or effective temperature
we would extract, depending on the model. Note however that the following two kinematic effects can also affect the interpretation of strangeness ratios: firstly, the energy cost of producing a strange-quark rest mass, as compared to say an up or down one, is relatively more significant for a hadronising system with small invariant mass than for one with large invariant mass; thus we expect small-mass systems to exhibit lower strangeness ratios simply due to phase-space restrictions. Secondly, in the presence of cuts on $p_\perp$, the modelling of $p_\perp$ spectra affects the ratio of the observed to total strange production. 
\item Baryon production probes a unique colour topology
and has clear experimental signatures. Unlike mesons, baryons do not appear naturally in the ``leading-colour'' approximation ($N_C\to\infty$), which is so heavily relied upon by MC event generators. This makes their production somewhat more difficult to model and interpret but also more interesting as a probe of aspects beyond leading $N_C$. The dependence on mass and strangeness can be studied in detail, especially via dedicated Particle Identification (PID) capabilities. For example,
a recent ALICE study managed to identify and study strongly
decaying excited states such as $\Sigma(1385)^{\pm}$ and $\Xi(1530)^0$
\cite{Abelev:2014qqa}.  
\end{itemize} 
Combined, strangeness, baryon number, and mass (or spin-multiplet) constitute a powerful 3-dimensional ladder along which to chart the behaviour of the underlying event. 
All of the major LHC experiments can provide sensitive measurements for key quantities such as multiplicities of charged tracks, \kshort, and 
$\Lambda$, while the more specialised PID capabilities of the ALICE and LHCb experiments permit reconstruction of several of the more challenging (e.g.~multi-strange and excited) states that will be needed to complete the picture. The production of each particle species can then be understood in the context of
the overall modelling. For instance, a measurement concluding that the production of kaons is low in a given model can have vastly
different physics implications, depending on whether the model
underpredicts the rates of all other particles as well. 

In this study, the most powerful model discrimination arises from taking the ratios of averaged identified-particle yields as described in sec. \ref{sec:observable}. This highlights the salient physics while minimising the effects from global scaling differences between the considered models. 

Though we focus on central rapidities in this paper, we round off by
noting that, at forward rapidities, the particle production is
increasingly sensitive to the hadronisation of the beam
remnants, and to diffraction. In the context of MPI models, the amount
of energy (both total and transverse) scattered into the forward direction is also
sensitive to the shape of Parton Distribution Functions (PDFs) at low $x$ (see e.g.\ \cite{Skands:2014pea}). Thus measurements at
forward rapidities (e.g.~AFP, ALFA, LHCb, LHCf, TOTEM) are uniquely
sensitive to these aspects and are required to understand the full
dynamics. An obvious example of interest is the colour structure of the beam
remnant, which is tied in with that of the MPI by overall colour
conservation\footnote{Since the colliding hadrons are colour neutral, the colour charge of each hadron remnant must be equal and opposite (in phase) to the sum of those of the MPI initiators (including the hardest-interaction initiators) taken from that hadron.}.
Studying the fragmentation process for diffractively
tagged events would also furnish insights complementary to those provided by the measurements we propose here. 
\section{Monte Carlo Models \label{sec:mc}}
For this study, we include the set of MC event generators and tunes listed in \tabRef{tab:mc}. This permits us to span a significant range of possible dynamic effects, including colour ropes (with \dipsy), hydrodynamics (with \epos), and colour reconnections (``CR'', with \py). The varying dynamical assumptions imply that these models will generate qualitatively different predictions for the scaling with transverse activity in one or more of the observables we consider below. In all cases, an inclusive `minimum bias' event sample was generated for analysis at $\sqrt{s} = 13 \TeV$. The key features of these models, in the context of this paper, are described below. 10 million events were simulated for each sample. The analysis was performed with \textsc{Rivet}~\cite{Buckley:2010ar}, including the use of \textsc{FastJet}~\cite{Cacciari:2011ma}.
\begin{table}[]
\centering
\caption{MC generators, versions and tunes used in this paper.}
\label{tab:mc}
\begin{tabular}{@{}lll@{}}
\toprule
 \multicolumn{2}{c}{Version} & \multicolumn{1}{c}{Tune} \\ \midrule
\dipsy                        & ThePEG++ 2015-08-11        & NoSwing            \\
\dipsy                        & ThePEG++ 2015-08-11        & Rope               \\
\epos                         & 1.3                        & LHC              \\
\py                           & 8.210                      & Monash           \\
\py                           & 8.210                      & Monash + new CR    \\ \bottomrule
\end{tabular}
\end{table}

\subsection{Pythia and colour re-connections}
The description of soft-inclusive physics in \pyNonum is rooted in
perturbation theory, starting from a resummed (unitarised) picture of 
multiple perturbative parton-parton interactions
(MPI)\cite{Sjostrand:1987su} supplemented by (interleaved)
$p_\perp$-ordered parton showers~\cite{Sjostrand:2004ef}. 
This builds up a partonic substructure
which can act as the starting point for the non-perturbative modelling of
each event\footnote{Exceptions are
  elastic and low-mass diffractive scatterings, which are modelled
  without a perturbative era.}. However, when there is more than one MPI
the perturbative era does not completely fix the detailed colour
structure of the beam remnant, nor  does it fix the colour correlations
between individual MPI systems~\cite{Sjostrand:2004pf}. Additionally, within
each system, ambiguities beyond leading colour could affect the formation of
strings, and strings could conceivably even interact
dynamically in the limited time between formation and fragmentation. 
Recent years have therefore seen increased explorations of new
alternatives though so far only a part of the full range of possibilities have
been explicitly addressed~\cite{Argyropoulos:2014zoa,Christiansen:2015yqa}.  

The default modelling of colour flow in \py~\cite{Corke:2010yf},
which is the one used by the baseline Monash
tune~\cite{Skands:2014pea}, is as follows: the partons of each MPI 
system are either allowed to form their own ``skeleton'' in colour space or, with
a user-modifiable probability, they are merged with the colour
structure of a higher-$p_\perp$ MPI system. In the latter case, each parton
from the lower-$p_\perp$ system is merged onto a string piece of the
higher-$p_\perp$ system where it will cause the least increase in total
string length. Within the context of this model, the
minimisation of string length is likely to be a physically reasonable dynamical principle, representing potential-energy minimisation, but
there is no a priori basis for guessing precisely what
reconnection-probability to choose, nor whether it should be constant at all CM energies. 

More recently, a model that attempts to anchor itself more firmly in
QCD was proposed~\cite{Christiansen:2015yqa}, which we here label ``Monash + New
CR'' (cf.~\tabRef{tab:mc}). It uses an approximation to the full 
group-theoretical weights from {\tt SU(3)} to compute the probabilities for alternative 
string topologies at the subleading-$N_C$ level, and is
again combined with a dynamical selection favouring topologies with
low string lengths. A novel feature of this model is that it allows for ``string junctions'' (colour-epsilon and -anti-epsilon structures) to form, which, combined with \pyNonum's model for junction fragmentation~\cite{Sjostrand:2002ip}, furnishes a new
source of baryon production. Moreover, since it is the global colour
structure of the event which gives rise to these additional 
baryons, their correlations (and net baryon-antibaryon compensation)
can happen over longer distances than is the case for the conventional (local)
baryon-production mechanisms in the string model.

In both cases, the fragmentation parameters of the string model are
constrained by LEP data~\cite{Skands:2014pea,Christiansen:2015yqa}. In particular,
this means that the strangeness fraction is essentially
fixed, modulo phase-space constraints and potential
  trigger-bias effects. The same is true for baryon production in the
  baseline CR model, while, in the New CR model, the baryon fraction can
 increase with the amount of CR. Exotic heavy baryons can also be
 formed, containing multiple charm or bottom quarks.

\subsection{DIPSY and rope hadronisation}
\dipsy \cite{Flensburg:2011kk} implements Mueller's dipole cascade model \cite{Mueller:1993rr} which operates in transverse coordinate space (unlike conventional showers which operate in momentum space). Instead of conventional PDFs, the model starts from an explicit representation of each proton being composed of three colour dipoles in impact-parameter space and rapidity. These are then evolved in rapidity space via iterated gluon emission, forming a dipole cascade. The resulting partonic final states are hadronised using the Lund string model, via an interface to \py~\cite{Sjostrand:2014zea}.

The recently developed rope extension to \dipsy~\cite{Bierlich:2014xba} allows geometrically nearby strings to act as a combined ``colour rope'', which can hadronise with a higher intrinsic tension. The consequences are larger $p_\perp$ kicks (relative to the string direction, which for soft-particle production in $pp$ collisions largely coincides with the $z$-axis) and production of more strange hadrons and baryons, the latter via probabilistic collapses of ropes to string junctions. The model also incorporates a mechanism called final-state ``swing'', which acts to minimise the masses of final-state colour-dipoles via colour reconnections, driven by {\tt SU(3)} colour rules combined with an ad-hoc probabilistic evolution kernel.  

To provide a reference case without these effects, we also include a ``NoSwing'' tune of \dipsy with parameters optimised without use of the final-state rope and swing mechanisms.

\subsection{EPOS and hydrodynamic core hadronisation}
The \epos MC \cite{Werner:2005jf} takes a Parton-Based Gribov Regge theory approach to event generation \cite{Drescher:2000ha}. String hadronisation in \epos is treated differently based on the local density of string segments per unit volume with respect to a critical-density parameter. Each string is classified as being in either a low density coronal region or in a high density core region. 

Corona hadronisation proceeds via unmodified string fragmentation whereas the core is subjected to a hydrodynamic evolution; i.e.\ it is hadronised including additional contributions from longitudinal and radial flow effects \cite{Werner:2007bf}. Core conditions are easily satisfied in ion collisions, however even for an average $pp$ collision $(N_{\rm ch} = 30, |\eta| < 2.4)$ at $\sqrt{s} = 7 \TeV$, around 30\% of central particle production arises from the core region. This rises to 75\% for $N_{\rm ch} = 100$. 

The \epos LHC \cite{Pierog:2013ria} tune considered here is a dedicated parameter set used to describe all LHC energies and incident particles. A loss of particle multiplicity due to radial flow rescaling is a feature needed to model ion collisions, however such effects are not observed in $pp$ collisions. To compensate, the LHC tune adds an additional parametrisation which modifies the radial flow of the small, dense cores typical of $pp$ interactions to fix the cluster mass and hence maintain the particle multiplicity. The total momentum is then rescaled after the radial boost to preserve energy conservation.   
\section{Observables \label{sec:observable}}
Fiducial cuts are applied to the MC generator output to approximate experimental sensitivity and this results in an inclusive set of particles formed of two components. The `prompt charged' component of the inclusive set consists of charged particles with $\pt > 200 \MeV$, $|\eta| < 2.5$, lifetime $\tau > 300$ ps and which are not created from the decay of a state with $30 < \tau < 300$ ps. This set is dominantly $\pi^\pm$, $K^\pm$, $p$ and $\bar{p}$. The definition is based around the ATLAS fiducial selection in ref. \cite{Aad:2016mok}. The second component consists of `identifiable prompt strange hadrons'; here both charged and neutral strange hadrons are included if they typically undergo weak decay to one or more charged particles. These states are also required to satisfy $\pt > 200 \MeV$, $|\eta| < 2.5$ and for themselves to not be created from the decay of other states with  $30 < \tau < 300$ ps\footnote{So for example $\Lambda$ originating from $\Xi^-$ decay are excluded.}. This set is comprised of \kshort, $\Lambda$, $\bar{\Lambda}$, $\Xi^\pm$, $\Sigma^\pm$, $\bar{\Sigma}^\pm$ and $\Omega^\pm$.

Track jets are clustered from prompt charged and prompt identifiable strange hadrons. They are reconstructed with the anti-$k_t$ algorithm \cite{Cacciari:2008gp} using radius parameter $R=0.4$, the leading jet is required to be within $|\eta| < 2.3$.

\subsection{Underlying Event Observables \label{sec:UEobs}}

\begin{figure}
\centering
\includegraphics[width=0.5\textwidth]{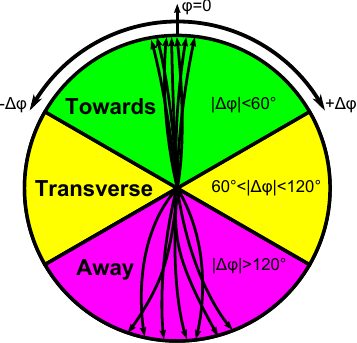}
\caption{Illustration of the {\em towards}, transverse and {\em away} regions of the azimuth with respect to the leading track jet in the event.}
\label{fig:UE_Diag}
\end{figure}

Standard Underlying Event nomenclature is used with respect to the leading track jet in the event as illustrated in \figRef{fig:UE_Diag}. In this note we will consider quantities constructed from particles only in the transverse region as this region is least affected by contributions from the leading  $2\rightarrow 2$ hard scatter.

The following variables are considered within the transverse region, these are common variables used to parametrise the underlying event as they probe the activity level and both the size and shape of the particle \pt\ distribution. The $\left<\right>$ notation implies an average over all events:
\begin{itemize}
\item $\left< \Nch \right>$: The event-averaged multiplicity of the inclusive set of particles (prompt charged plus prompt identifiable strange hadrons).
\item $\left< N(X) \right>$: The event-averaged multiplicity of identified particle(s), $X$.
\item $\left< \sum \pt \right>$: The event-averaged scalar sum of the transverse momentum of the inclusive set of particles.
\item $\left< {\rm mean}\;\pt \right>$: The event-averaged mean \pt of the inclusive set of particles. 
\item $\left< {\rm RMS} \right>$: The event-averaged root mean square \pt of the inclusive set of particles, defined here for a given multiplicity, $\Nch$, as $\left< {\rm RMS} \right> = \sqrt{ \left< \sum \pt^2 \right>/\Nch }$.
\end{itemize}
\subsection{Relative Transverse Activity Classifier, $\Rt$ \label{sec:rt}}
Above a lower threshold  corresponding to the onset of the UE plateau in the transverse region (roughly, $p_{\perp\mrm{lead}} > 10 \GeV$), the mean values of UE quantities calculated over the full inclusive set of particles have little dependence on the leading track-jet \pt: the multiple soft scatters which contribute the majority of the UE are largely independent of the leading jet. The slow rise of the UE plateau in the transverse region \emph{is} understood to be due to additional contributions from wide-angle radiation associated with the hard scatter, but this effect becomes significant only for jet $\pt > 50 \GeV$ \cite{Aad:2014hia}. In this study, we focus on events just above the onset of the plateau, with $10 \GeV < p_{\perp\mrm{lead}} < 30 \GeV$.

\begin{figure}[tp]
        \begin{subfigure}[b]{0.5\textwidth}
                \centering
                \includegraphics[width=.98\linewidth]{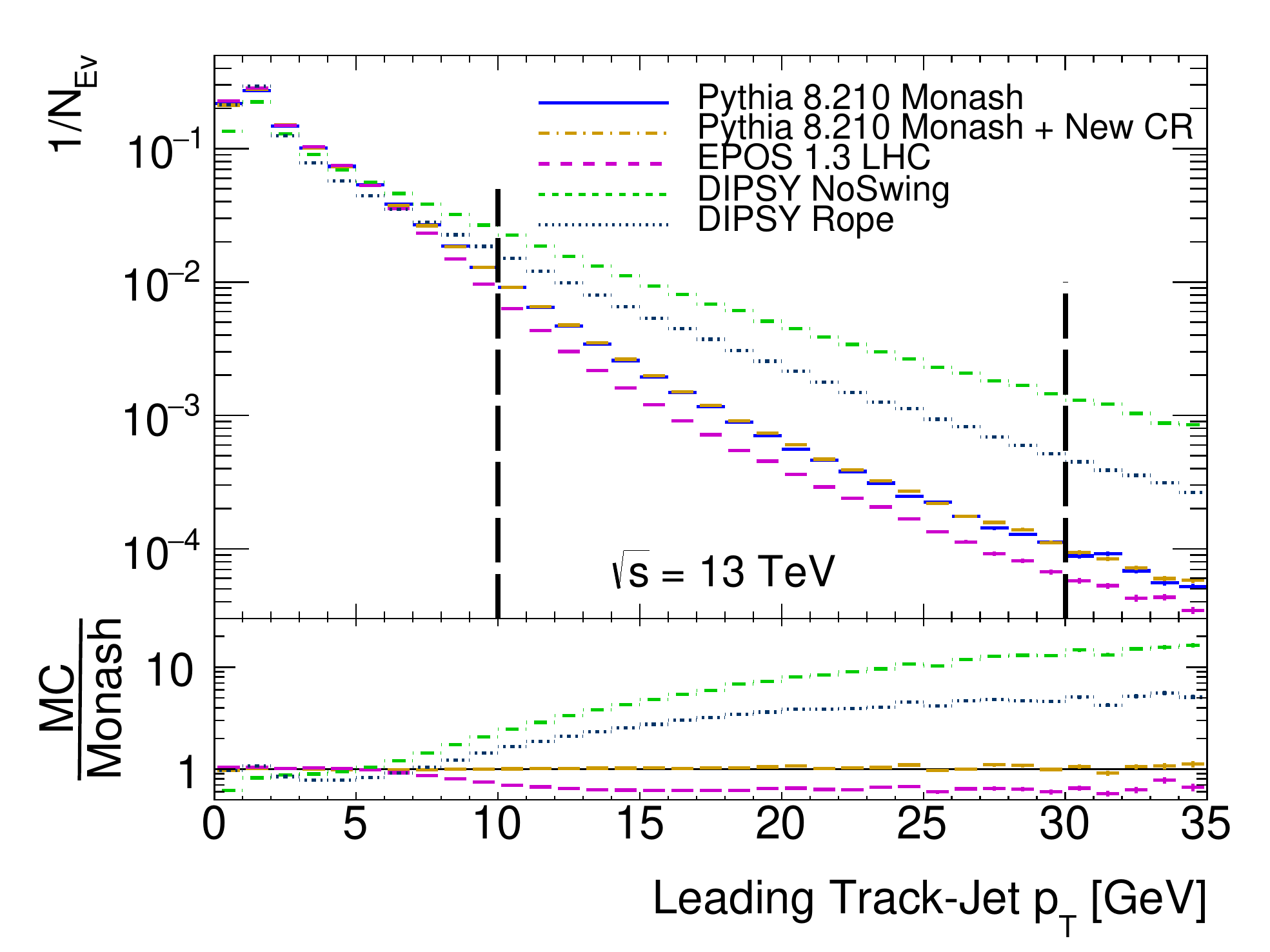}
                \caption{}
                \label{fig:ue-demo:a}        
        \end{subfigure}%
        \begin{subfigure}[b]{0.5\textwidth}
                \centering
                \includegraphics[width=.98\linewidth]{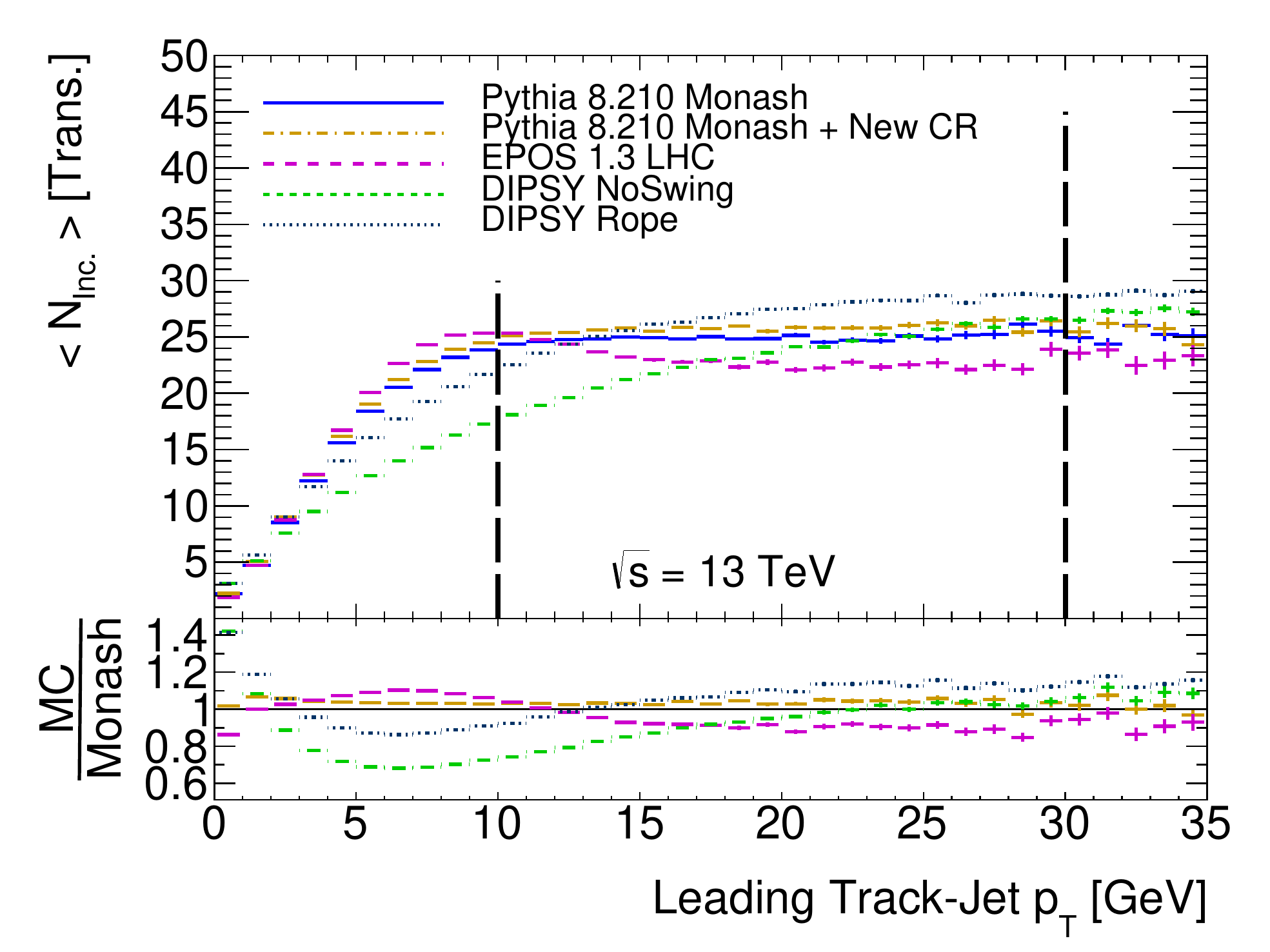} 
                \caption{}
                \label{fig:ue-demo:b}
        \end{subfigure}%
        \caption{Leading track-jet \pt spectrum, normalised to unity (a) and $\left<\Nch\right>$ in the transverse region as a function of leading track-jet \pt. Dashed vertical lines indicate the \pt range used in this paper. }
        \label{fig:ue-demo}
\end{figure}

The shape of the leading track-jet \pt\ spectrum in the various MC generators considered is shown in \figRef{fig:ue-demo:a}. Dashed vertical lines indicate the lead $p_{\perp}$ range used in this study. The \dipsy tunes clearly exhibit a significantly harder jet \pt\ spectrum than \py\ or \epos. However, when we consider the average inclusive-particle yields in the transverse region, shown in \figRef{fig:ue-demo:b}, we observe that the activity in the transverse region is roughly constant and quite comparable between the models, independently of the jet \pt\ spectrum. We therefore stress that any mis-modelling in the absolute rate of jet events is expected to have little effect on the results in this study due to the profiled mean normalisation. Note that the most extreme variant of \dipsynoswing does exhibit a delayed onset of the $\left< \Nch \right>$ plateau. In the context of an experimental analysis, the robustness of the conclusions could be explicitly validated by comparing with one or more higher jet $p_\perp$ windows, say 20-40 \GeV\ and/or 30-50 \GeV, statistics allowing. 

\begin{figure}[tp]
        \centering
        \includegraphics[width=.7\linewidth]{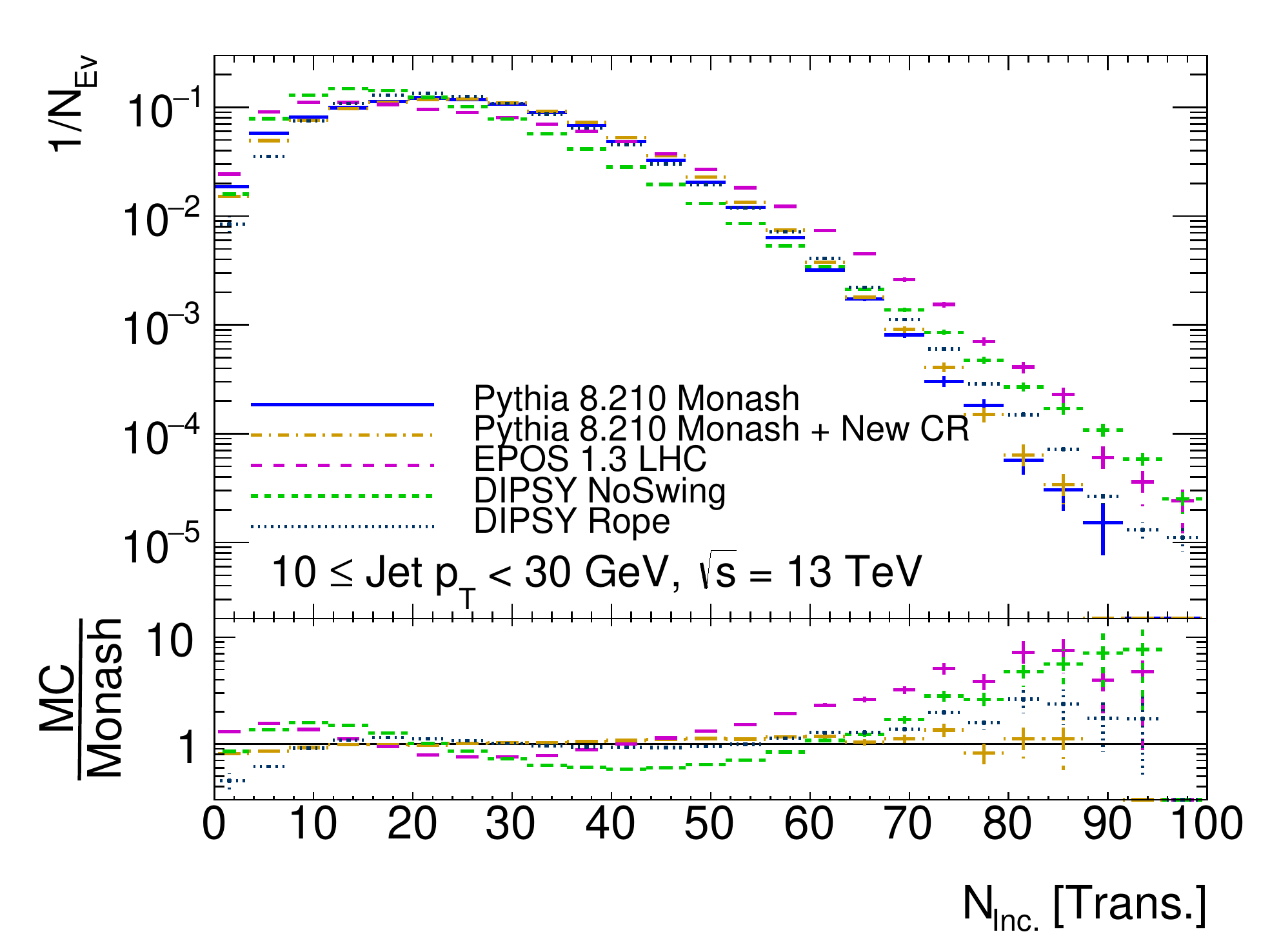}
        \caption{The transverse $\Nch$ distribution for events with leading track jet in the range $10~\leq~\pt~30~<\GeV$, normalised to unity. Colour online.}
        \label{fig:ue-nch}
\end{figure}

Although average quantities like $\left<\Nch\right>$ in \figRef{fig:ue-demo:b} do not vary considerably with \pt over this range of jet energies, the distributions underlying these mean values can be very broad. The full probability distribution of \Nch\ is shown in \figRef{fig:ue-nch}, integrated over all events with leading track-jet $p_\perp$ values in the range 10 -- 30 \GeV. Such very broad distributions with tails much wider than Gaussians ($\Gamma \gg \sqrt{\left<\Nch\right>}$) are typical in minimum-bias and underlying-event studies. They imply that there is a very large dynamic range between transverse-region activities significantly larger or smaller than the mean. Given that the approximately constant average UE level is by now well established, we believe that these extremes, which are typically hidden in studies of `average' UE properties, are the next natural focus of study. 

In order to obtain an axis which allows for the investigation of the modelling of proton interactions as a function of the event activity, we reclassify events with a leading track jet in the range $10 \leq \pt < 30 \GeV$ based on  their per-event transverse activity with respect to the mean:
\begin{equation} \label{eq:ac}
  \Rt = \frac{\Nch}{\left< \Nch \right>}.
\end{equation}
We find this normalisation choice (which in minimum-bias contexts is referred to as the KNO variable~\cite{Koba:1972ng}) to be useful since $\Rt=1$ then cleanly divides events with ``higher-than-average'' UE from ``lower-than-average'' ones, irrespective of CM energy or applied cuts. We note however that an absolute normalisation would be the preferred choice for determining, e.g., whether events with a fixed number of particles behave the same at all CM energies. 

For each MC model, the value of the denominator in \eqRef{eq:ac} corresponds to the mean values of the distributions in \figRef{fig:ue-nch}, which are tabulated in \tabRef{tab:denominator}. All models predict a mean transverse multiplicity in the range 21--26 and a width of around 13 (where Poissonian fluctuations would predict a much smaller width, $\sim \sqrt{25} = 5$).
\begin{table}[t]
\centering
\caption{The average fiducial $\left< \Nch \right>$, the denominator of \Rt, for the MC models considered along with the width of the distributions.}
\label{tab:denominator}
\begin{tabular}{@{}llll@{}}
\toprule
Generator & Tune & $\left< \Nch \right>$  & $\sigma$ \\ \midrule
\py         & Monash                   & 24.7   &  12.5    \\
\py         & Monash + New CR          & 25.5   &  12.6    \\
\epos       & LHC                      & 24.2   &  14.6    \\
\dipsy      & NoSwing                  & 21.3   &  12.2    \\
\dipsy      & Rope                     & 25.1   &  12.0    \\
\bottomrule
\end{tabular}
\end{table}

Measuring UE quantities versus \Rt yields sensitivity to rare events with exceptionally large or small transverse activity with respect to the average event. The lower requirement on the leading track-jet \pt acts to suppress soft-periphery and diffractive interactions by ensuring that a hard scatter was present while the upper requirement limits the contamination by wide-angle radiation off the hard scatter which increases slowly with the \pt\ of the leading jet \cite{Aad:2014hia}.

In the context of MPI models, the case where only a small amount of energy is deposited in the transverse region implies that only a small number of MPI occurred in that event. This therefore affords an opportunity to measure event properties in an `MPI-suppressed' environment, where fragmentation properties may be closer to those of $e^+e^-$ collisions than the average $pp$ jet event. 

In the other extreme where the activity in an event is many times larger than the mean, new dynamic effects may become significant. Some different models which may give rise to modified behaviour in dense $pp$ interactions were discussed in sec. \ref{sec:mc}. Such modifications may be important in describing effects already observed in data such as \cite{CMS:2012qk,Aad:2015gqa} and \cite{Chatrchyan:2013qsa}, and it is important to develop further probes that shed light on how these effects develop as we move from dilute to dense environments. 
\begin{figure}[t]
 \begin{subfigure}[t]{0.5\textwidth}
                \centering
                \includegraphics[width=.985\linewidth]{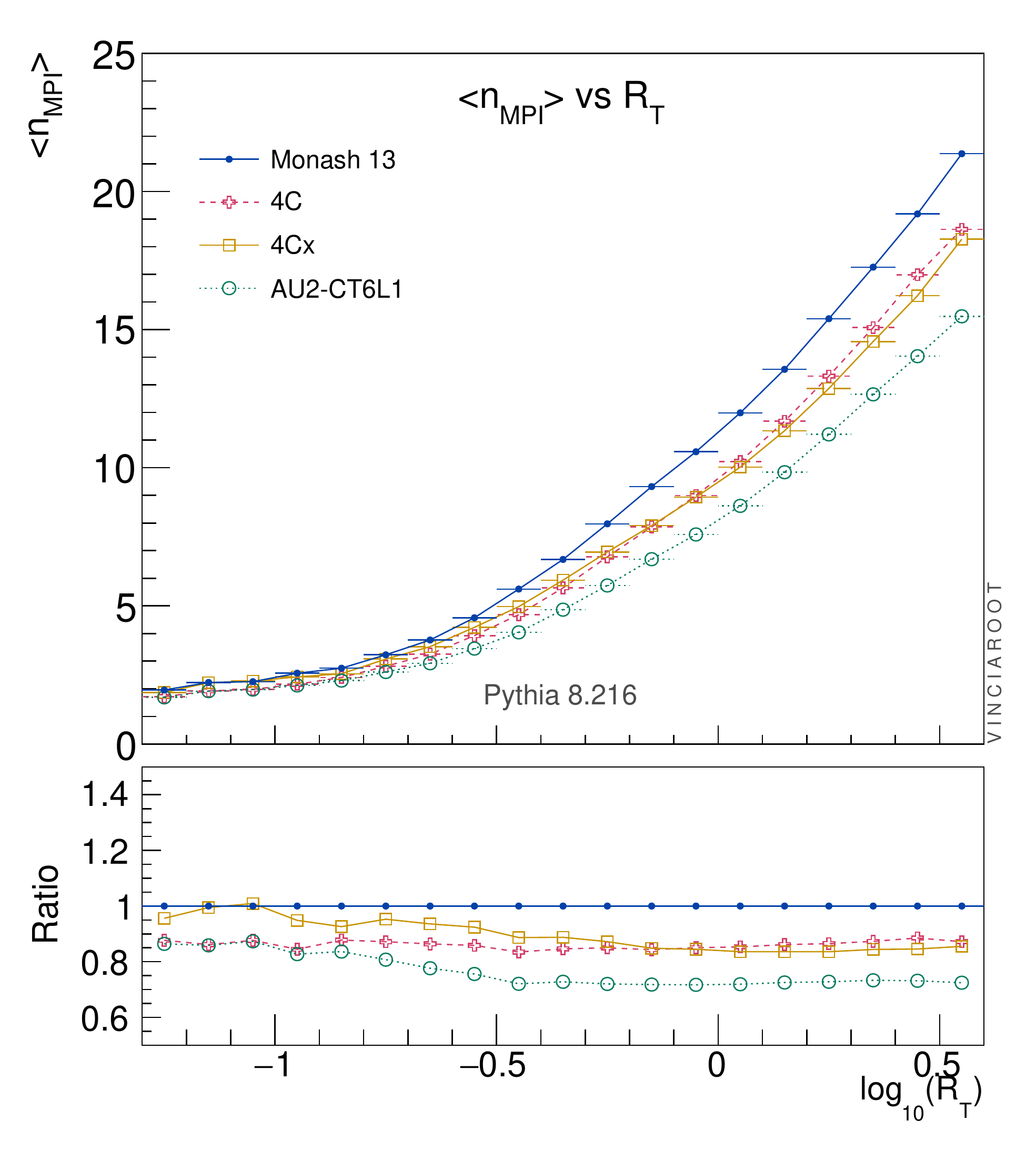}
                \caption{}
                \label{fig:nMPIvsAT}
        \end{subfigure}%
 \begin{subfigure}[t]{0.5\textwidth}
                \centering
                \includegraphics[width=.985\linewidth]{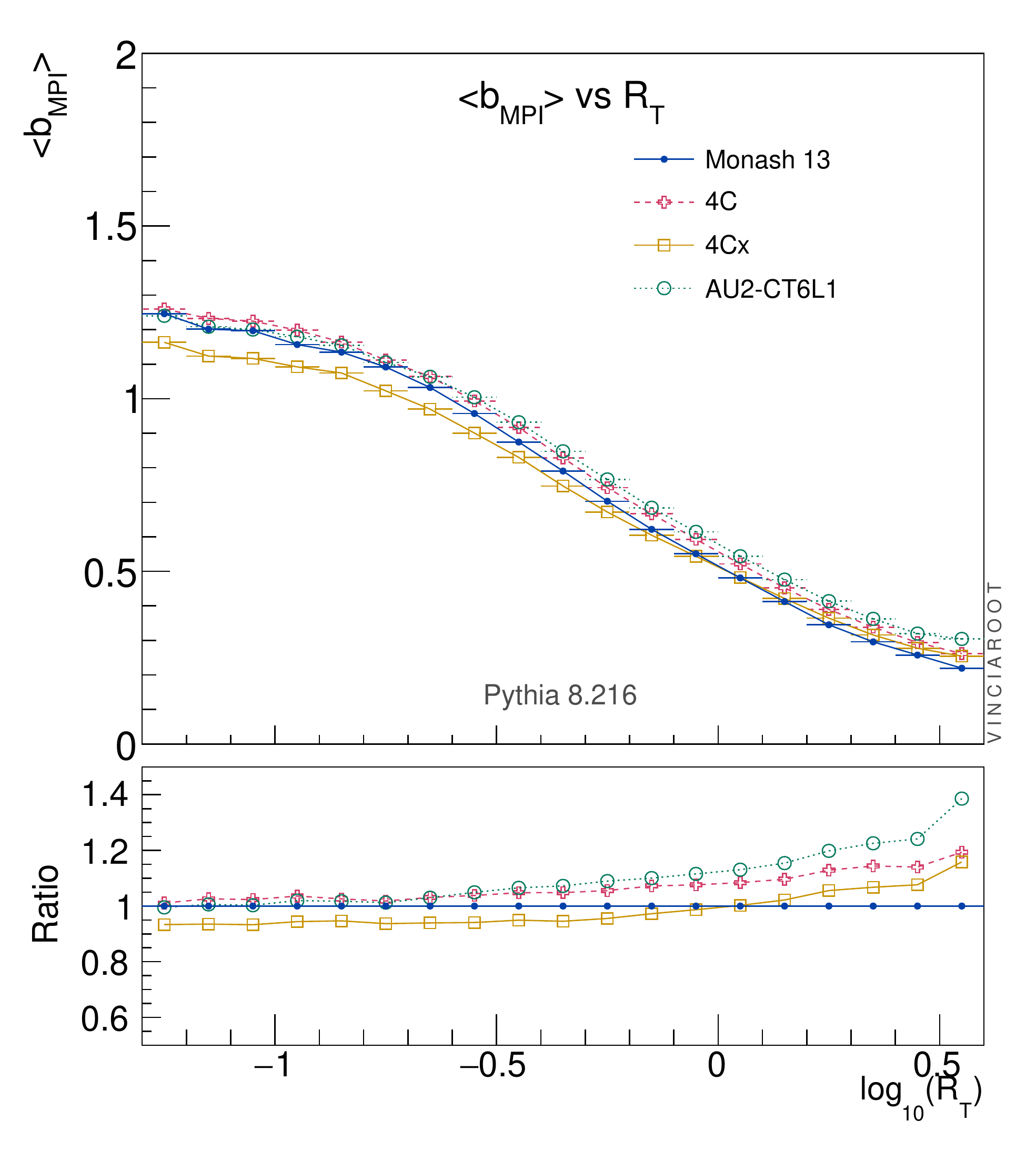}
                \caption{}
                \label{fig:bMPIvsAT}
        \end{subfigure}%
        \caption{The dependence of (a) the average number of parton-parton interactions (including the hardest interaction, hence $\left<n_\text{MPI}\right>\ge 1$) and (b) the average hadron-hadron impact parameter versus the transverse activity fraction, \Rt, defined in the text, for a representative set of \py tunes, for $pp$ collisions at 13 \TeV\ with at least one track jet (defined in text). Colour online.}
        \label{fig:nbMPI}
\end{figure}

Illustrated in \figRef{fig:nbMPI} are the average number of parton-parton interactions (MPI) and the average transverse proton-proton impact parameter $\left<b_\mrm{MPI}\right>$, with the latter normalised such that 1 corresponds to the impact parameter of an average minimum-bias event. Both are plotted as a function of $\log_{10}(\Rt)$ for four different tunes of \py. For the most active events ($\log_{10}(\Rt)=0.5$ corresponding to $\Rt=\sqrt{10}\sim3.2$ times higher-than-average UE activity) the average number of MPI increases by roughly a factor 2 relative to the mean (at $\log_{10}(\Rt)=0$), and the events are roughly twice as central as the average in this jet $p_\perp$ window (which in turn are twice as central as the average minimum-bias event). For low-activity events, with less than a tenth of the average UE activity, $\log_{10}(\Rt)<-1$, an average of less than 2 MPI per event are predicted by these models, with an average impact parameter even larger than for minimum-bias events, $\left<b_\mrm{MPI} \right> > 1$ for $\log_{10}(\Rt)<-0.5$. According to these models, our axis thus allows us to scan over almost an order of magnitude in both the average number of MPI and the average impact parameter, for fixed jet $p_\perp$.

For completeness, we note that relative UE activity could also have been classified using summed transverse-region \pt\ or by using the jet median/area techniques referred to in the introduction. Our choice of $\Nch$ was based on two factors: firstly, we have a very direct relation with charged-particle multiplicity based minimum-bias studies of similar quantities, and secondly the fact that heavier hadrons exhibit harder \pt\ spectra can lead to undesirable biases in a classifier based on $\sum \pt$. For example a low $\sum \pt$ UE would be biased to be made up entirely of pions since these have the softest spectra. Such biases would  complicate the interpretation of effects we wish to study such as any evolution in total strangeness and baryon fractions. 
\section{Inclusive Particle Set Results \label{sec:incresults}}
\begin{figure}[tp]
        \begin{subfigure}[b]{0.5\textwidth}
                \centering
                \includegraphics[width=.99\linewidth]{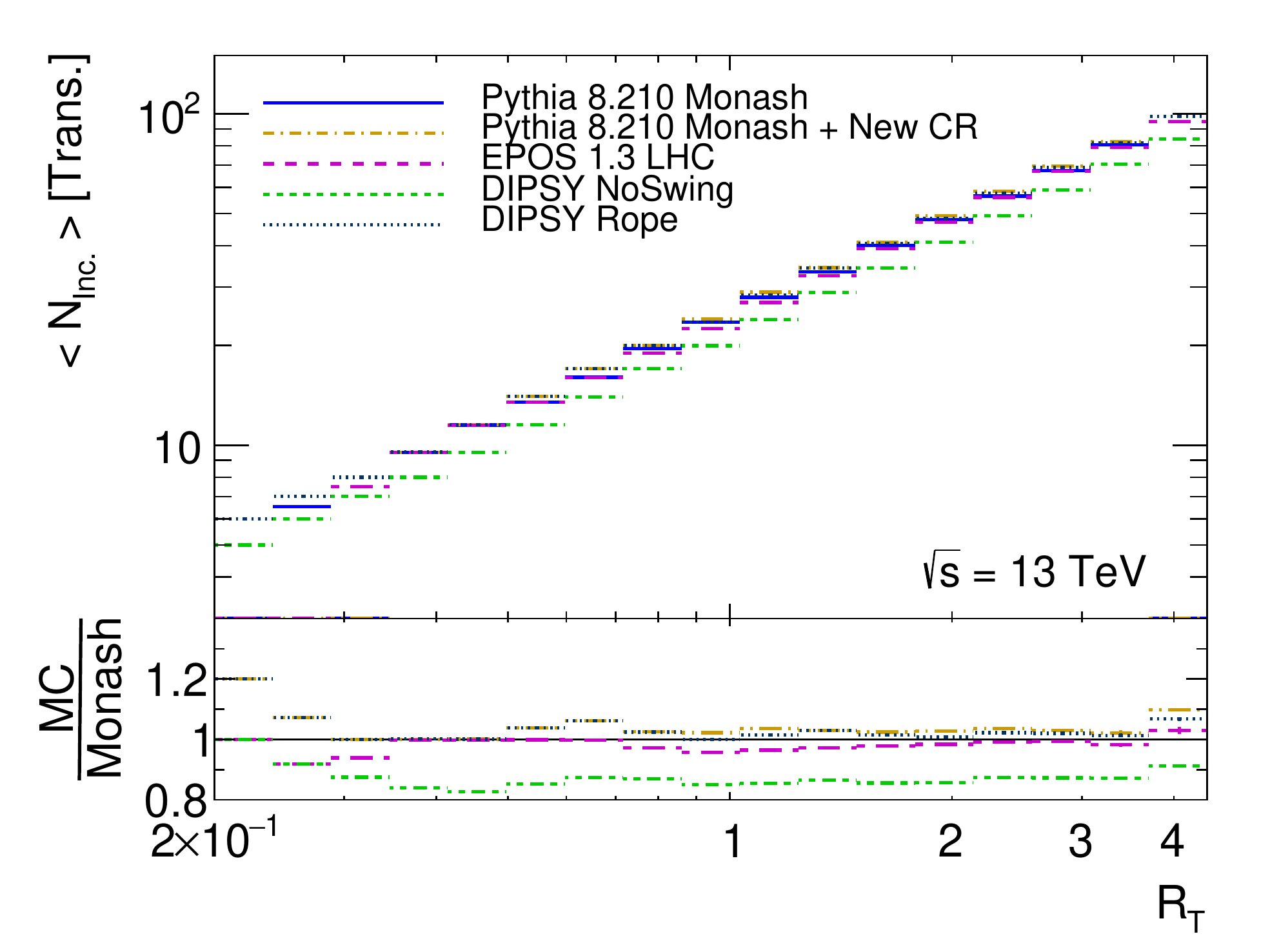}
                \caption{}
                \label{fig:ue-ac:a}
        \end{subfigure}%
        \begin{subfigure}[b]{0.5\textwidth}
                \centering
                \includegraphics[width=.99\linewidth]{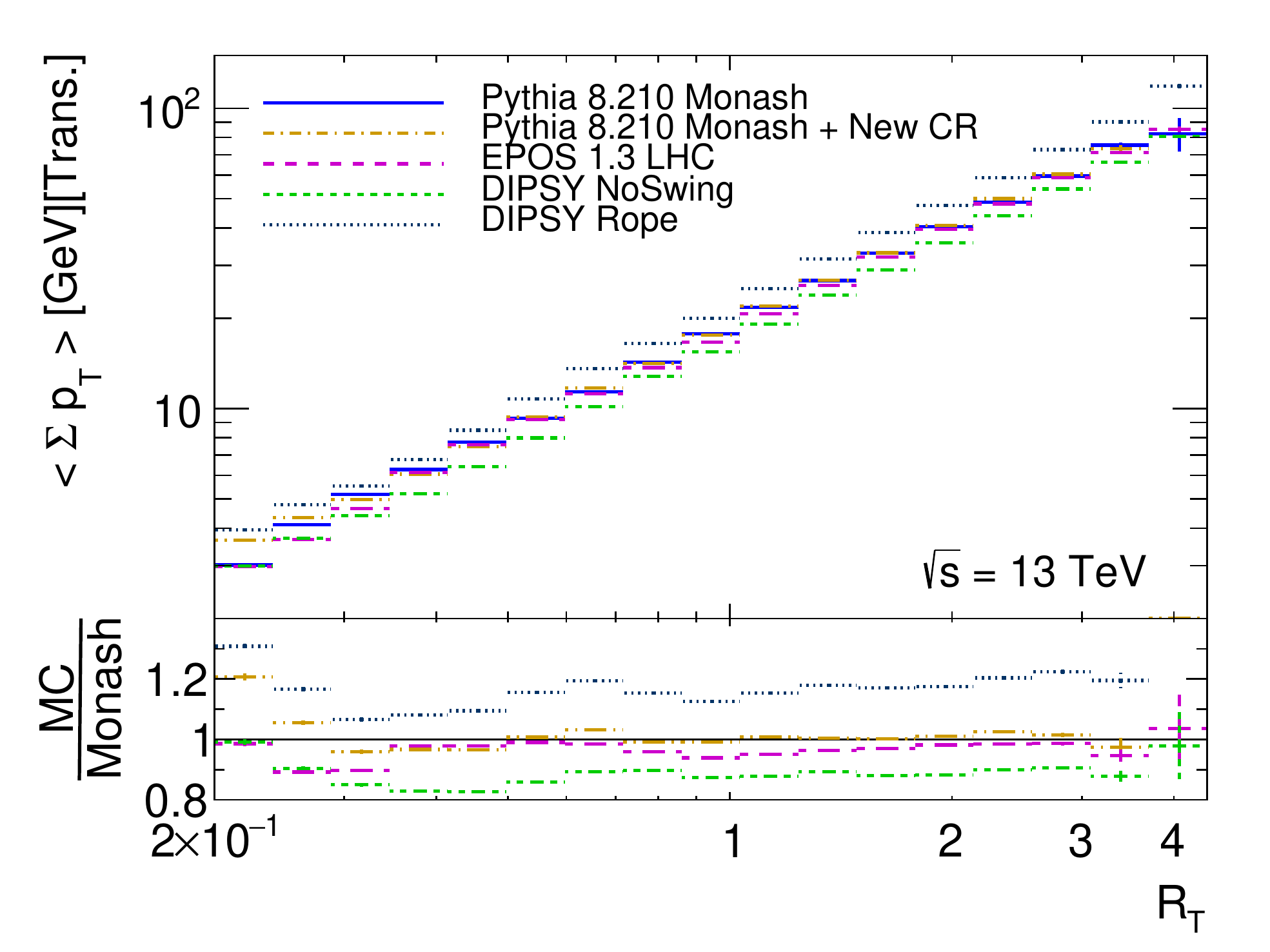}
                \caption{}
                \label{fig:ue-ac:b}
        \end{subfigure}%
        \caption{Average multiplicity (a) of the set of inclusive particles and sum \pt (b) as a function of the per-event activity classifier \Rt. All ratios are relative to \py.210 Monash. Colour online. }
        \label{fig:ue-ac}
\end{figure}
As a precursor to the main study of identified-particle ratios, we first consider the evolution of the inclusive particle spectra with \Rt. In \figRef{fig:ue-ac}, MC predictions for the average \Nch and $\sum \pt$ of the inclusive set of particles are plotted as a function of \Rt.
\begin{figure}[tp]
        \begin{subfigure}[b]{0.5\textwidth}
                \centering
                \includegraphics[width=.99\linewidth]{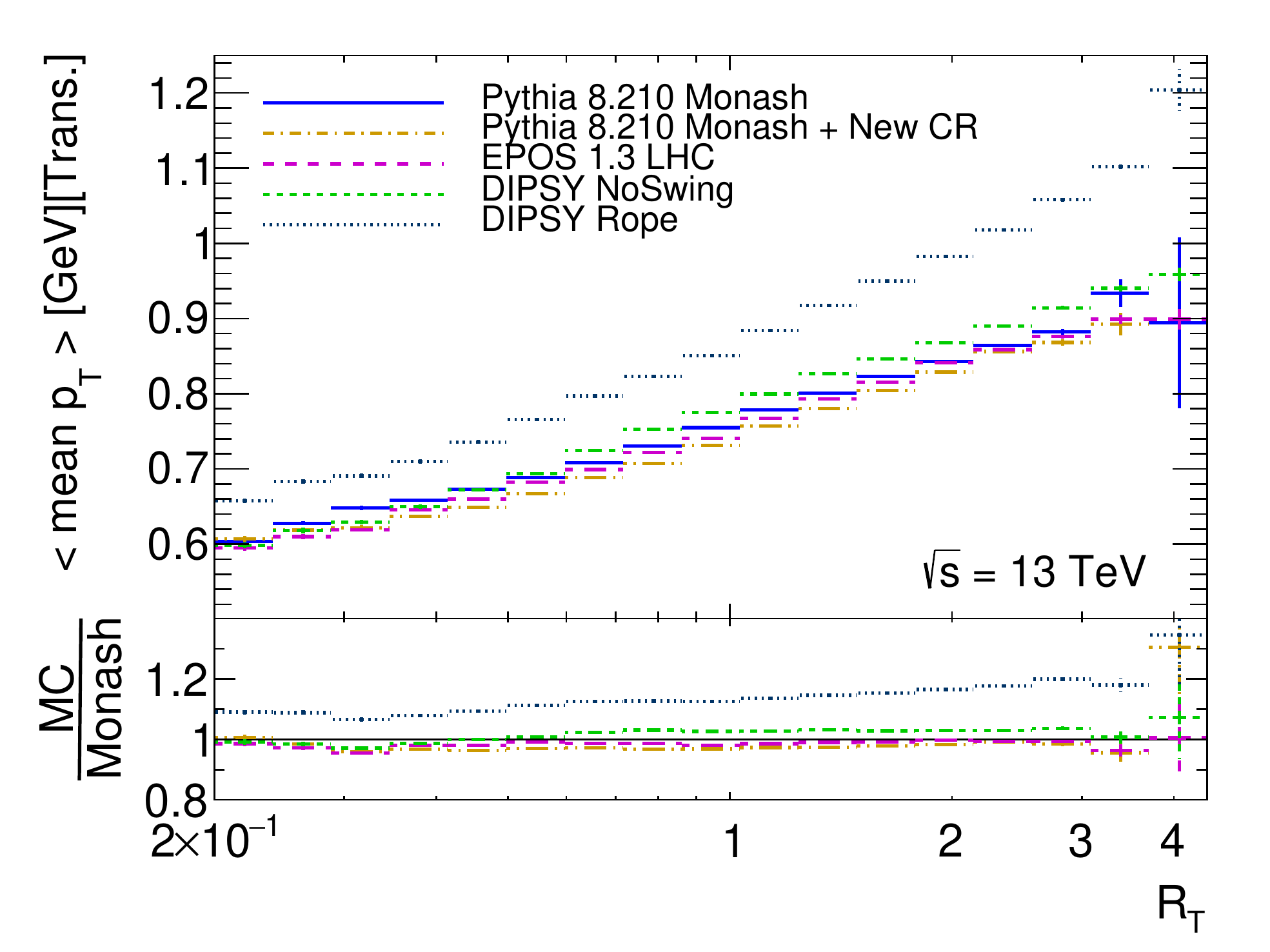}
                \caption{}
                \label{fig:mean-ue:a}
        \end{subfigure}%
        \begin{subfigure}[b]{0.5\textwidth}
                \centering
                \includegraphics[width=.99\linewidth]{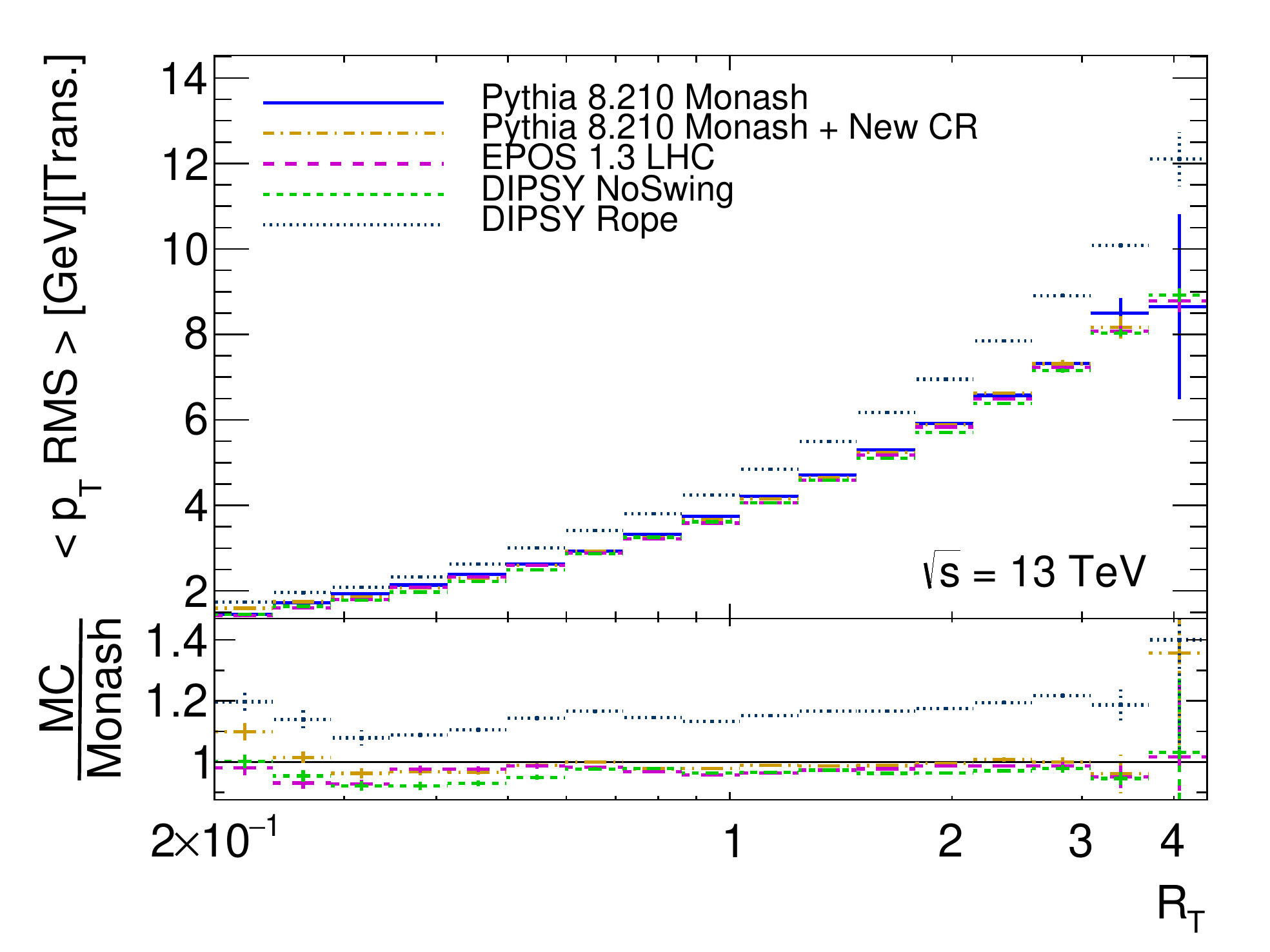}
                \caption{}
                \label{fig:mean-ue:b}
        \end{subfigure}%
        \caption{Average mean (a) and RMS (b) of the \pt distribution of the set of inclusive particles. All ratios are relative to  \py.210 Monash. Colour online. }
        \label{fig:mean-ue}
\end{figure}
The averages and widths of the inclusive \pt spectra are investigated in \figRef{fig:mean-ue}. The \dipsyrope tune generates events with the highest overall average energy density followed by \epos and \py. For \dipsyrope this is observed to be due to a similar average particle multiplicity as for \py combined with a 20\% harder mean \pt distribution and a similarly larger RMS. The \dipsynoswing tune has both the lowest average multiplicity spectra and energy density of the models, its mean \pt is similar to that of \py.

Only minor differences are observed for \py with the new colour-reconnection model with respect to the standard Monash tune in these distributions.
\section{Identified Particle Results \label{sec:identifiedresults}}
The production of identified particles is useful to investigate the evolution of the underlying event as a function of transverse activity levels in the event. The strange and baryon contents of the final state arising from non-perturbative effects are expected to be particularly sensitive to the modeling described in \secRef{sec:mc}.

\begin{figure}[tp]
    \begin{subfigure}[b]{0.5\textwidth}
            \centering
            \includegraphics[width=.99\linewidth]{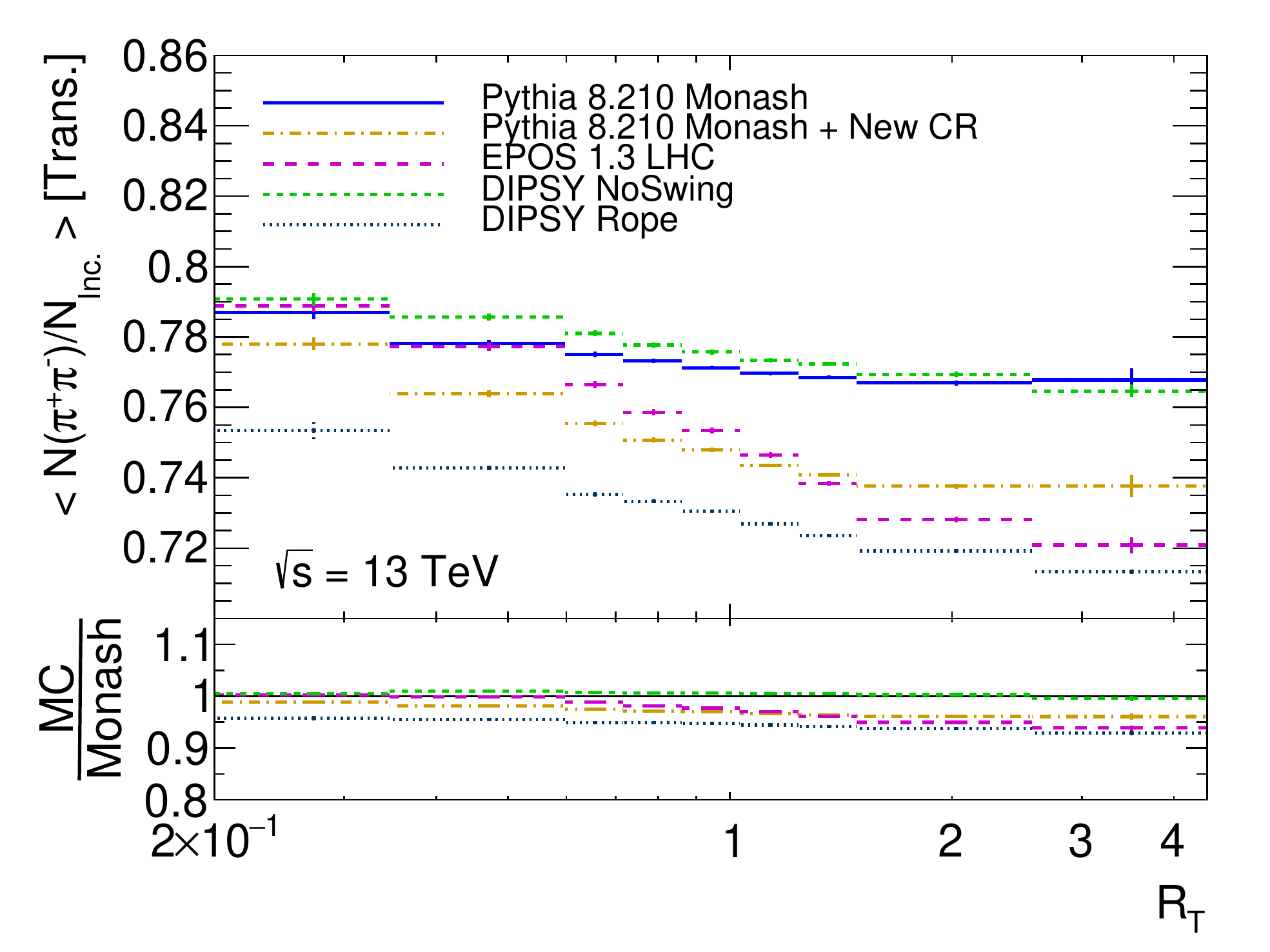}
            \caption{}
            \label{fig:identified-ueM:a}
    \end{subfigure}%
    \begin{subfigure}[b]{0.5\textwidth}
            \centering
            \includegraphics[width=.99\linewidth]{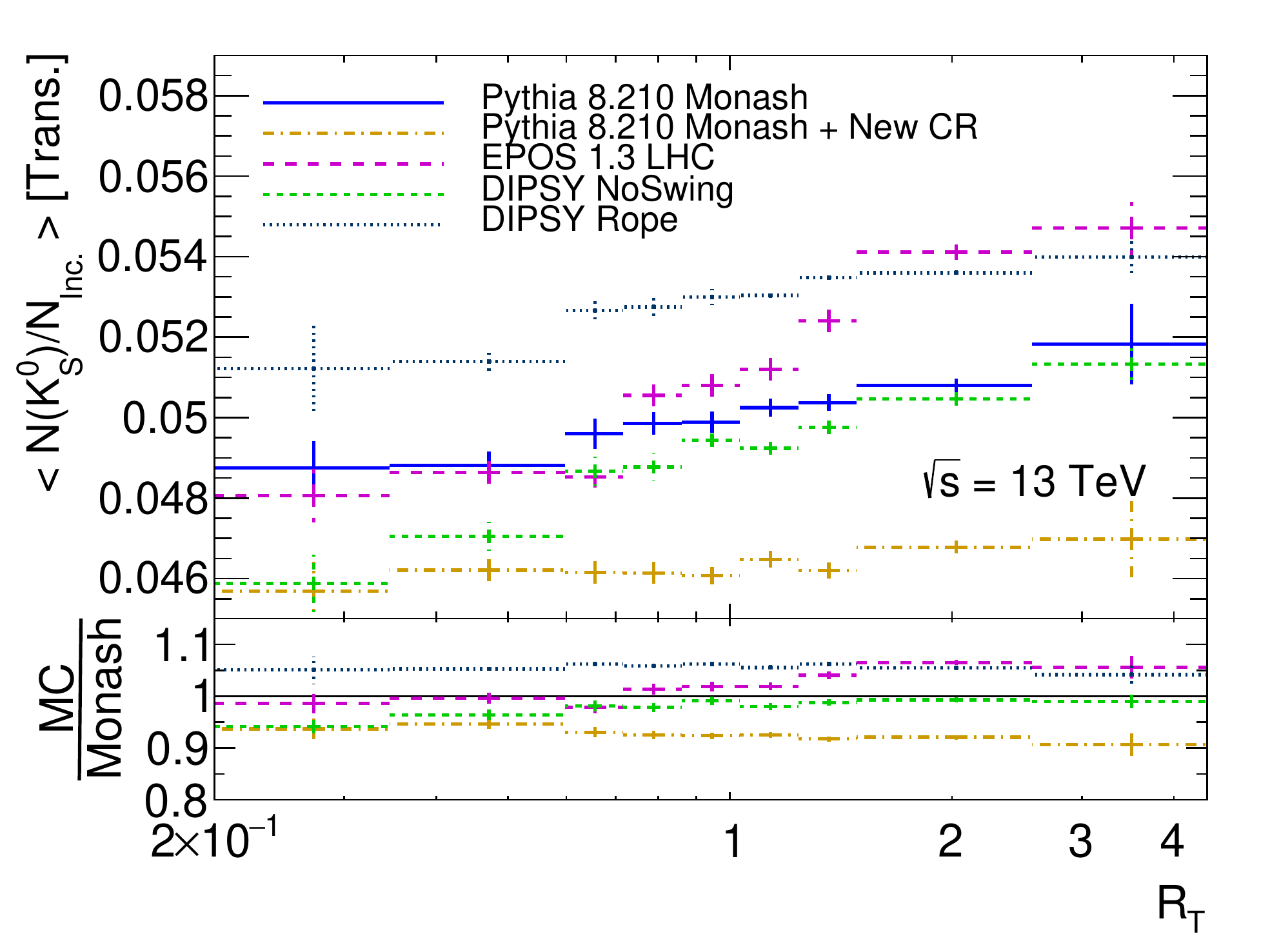}
            \caption{}
            \label{fig:identified-ueM:b}
    \end{subfigure}%
    \newline
    \begin{subfigure}[b]{0.5\textwidth}
            \centering
            \includegraphics[width=.99\linewidth]{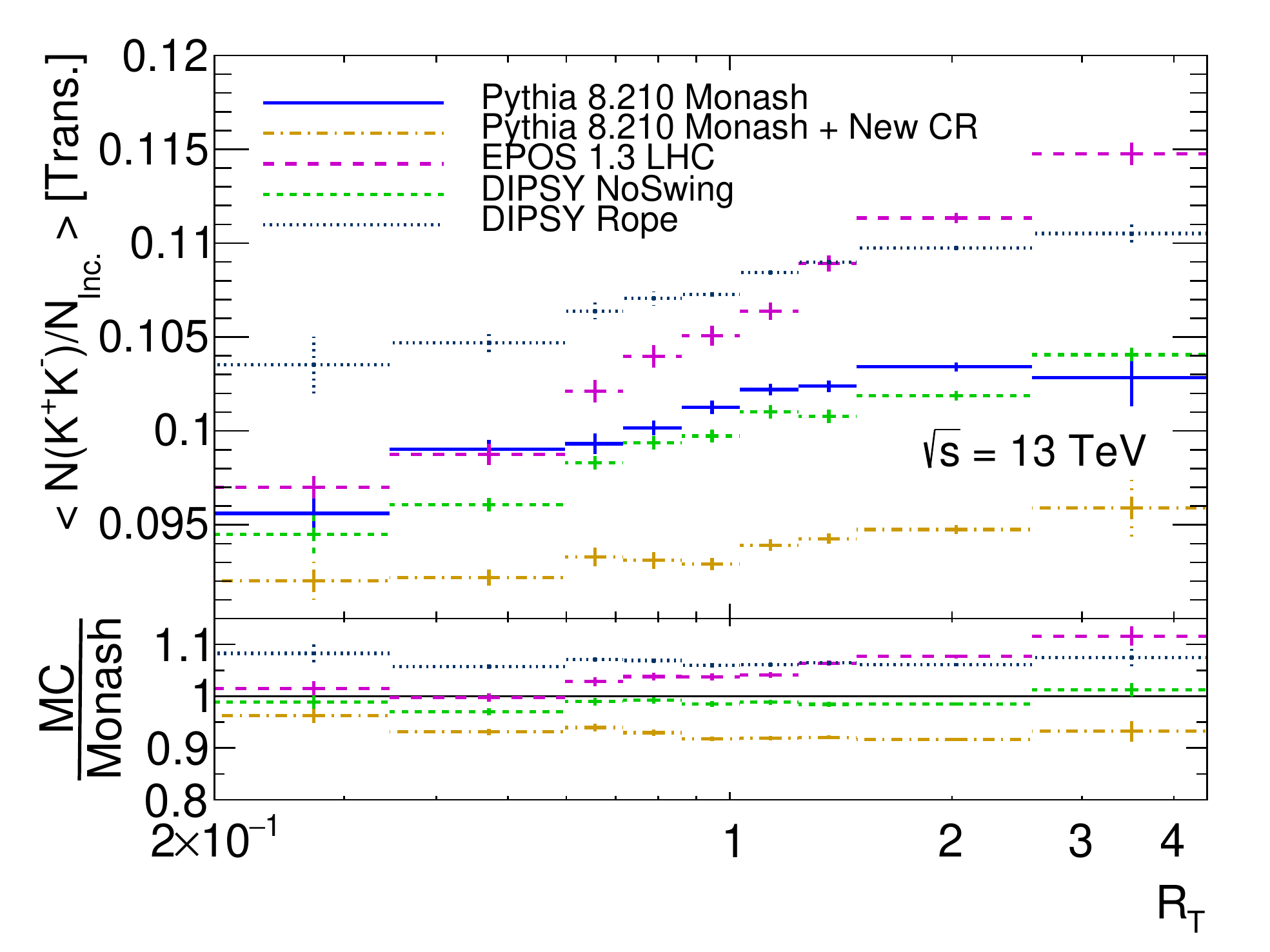}
            \caption{}
            \label{fig:identified-ueM:c}
    \end{subfigure}%
    \begin{subfigure}[b]{0.5\textwidth}
            \centering
            \includegraphics[width=.99\linewidth]{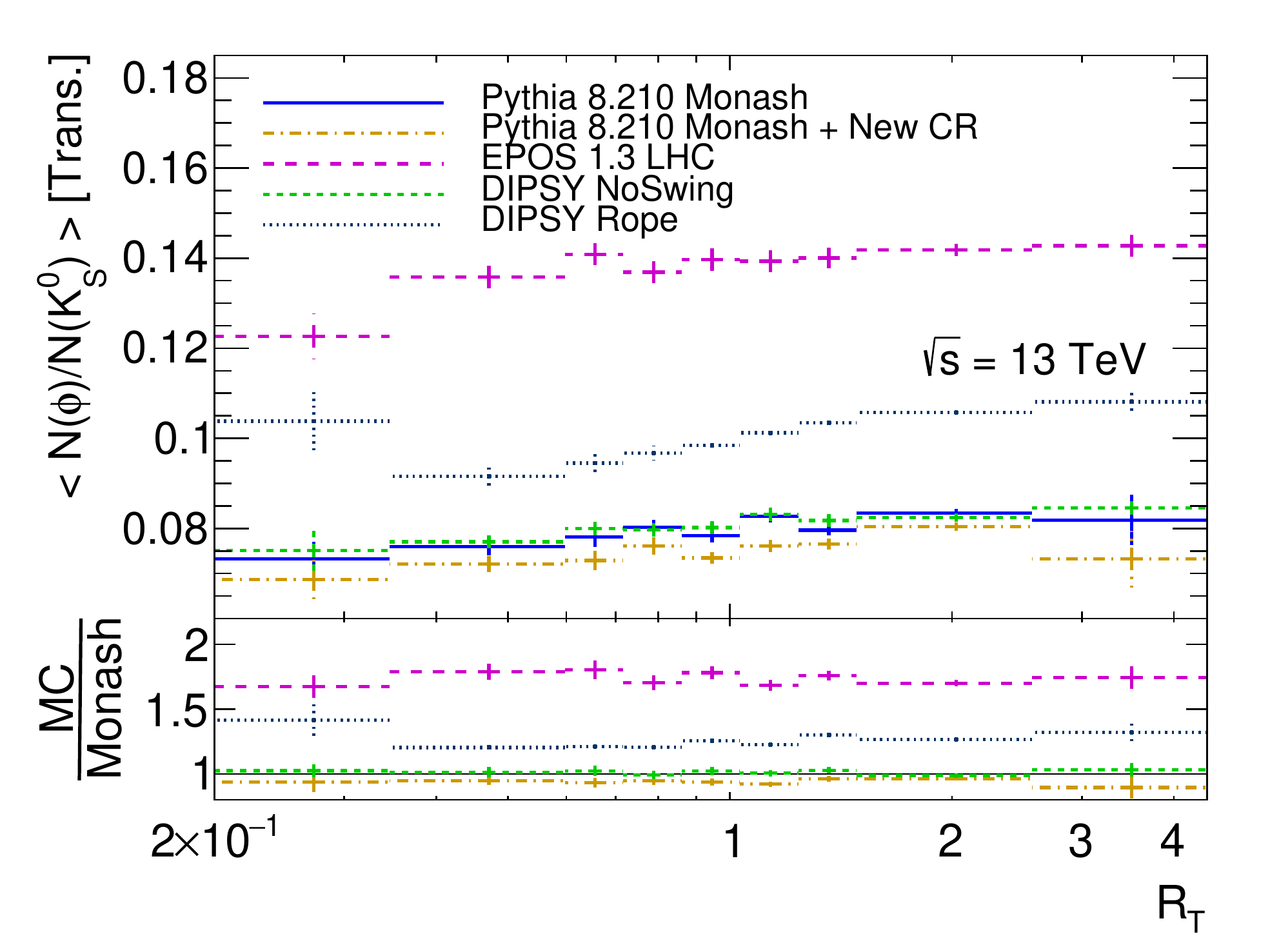}
            \caption{}
            \label{fig:identified-ueM:d}
    \end{subfigure}%
    \caption{Normalised average identified-meson yields in the transverse region as a function of \Rt for $\pi^+\pi^-/\Nch$ (a),  $\kshort/\Nch$ (b), $K^+K^-/\Nch$ (c) and $\phi/\kshort$ (d). Shown for different MC models and tunes. 
    All ratios are relative to  \py.210 Monash. Colour online. }
    \label{fig:identified-ueM}
\end{figure}

We normalise distributions to factor out any contribution from  differences in the overall multiplicity spectra of the generators and focus on ratios of particle yields, with the total yield of inclusive particles discussed in the previous section. Crucially, this also highlights any changes in the \emph{relative} suppression of identified particles, with respect to the inclusive sum. The meson fractions are plotted in \figRef{fig:identified-ueM}. The ratio of charged pions to inclusive particles is given first as the latter is used subsequently for normalisation as it is experimentally easier to access. The $\pi^\pm$ fraction is observed to fall from 76--79\% at low \Rt to 72--77\% at high \Rt, with the lowest fractions in \epos and \dipsyrope at high \Rt\ as expected due to increased strange and baryon production in these models as a function of \Rt. \py Monash + New CR, which does not incorporate strangeness enhancement but does allow for baryon enhancement, also exhibits a statistically significant drop.

The average \kshort and $K^\pm$ multiplicities with respect to the inclusive multiplicity are shown in \figRef{fig:identified-ueM:b} and \figRef{fig:identified-ueM:c}. A baseline increase in the kaon fraction with \Rt due to phase space effects is observed for the \py Monash and \dipsynoswing tunes. \epos predicts a significant strange meson enhancement at high \Rt whereas  \dipsyrope predict an enhancement for all \Rt, in both cases it reaches 5--10\%.  Interestingly \py Monash + new CR exhibits slightly lower charged- and neutral- kaon fractions at all \Rt. This is likely to be due to a smaller baseline strangeness fraction in that tune combined with a tendency of the model to create a large number of low-mass string systems in which the production of high-mass hadrons (including strange ones) is suppressed due to phase-space restrictions.

When the ratio of the doubly-strange $\phi(s\bar{s})$ meson multiplicity is taken with respect to that of the singly-strange \kshort in \figRef{fig:identified-ueM:d}, the new CR model in \py does not produce additional $\phi$ and is also in agreement with \dipsynoswing. Both \dipsyrope and \epos however, which both incorporate explicit strangeness enhancement, are shown to produce considerably more $\phi$ on average, up to 30\% for \dipsyrope and 75\% for \epos with respect to \py.
\begin{figure}[tp]
    \centering
    \begin{subfigure}[b]{0.5\textwidth}
            \centering
            \includegraphics[width=.99\linewidth]{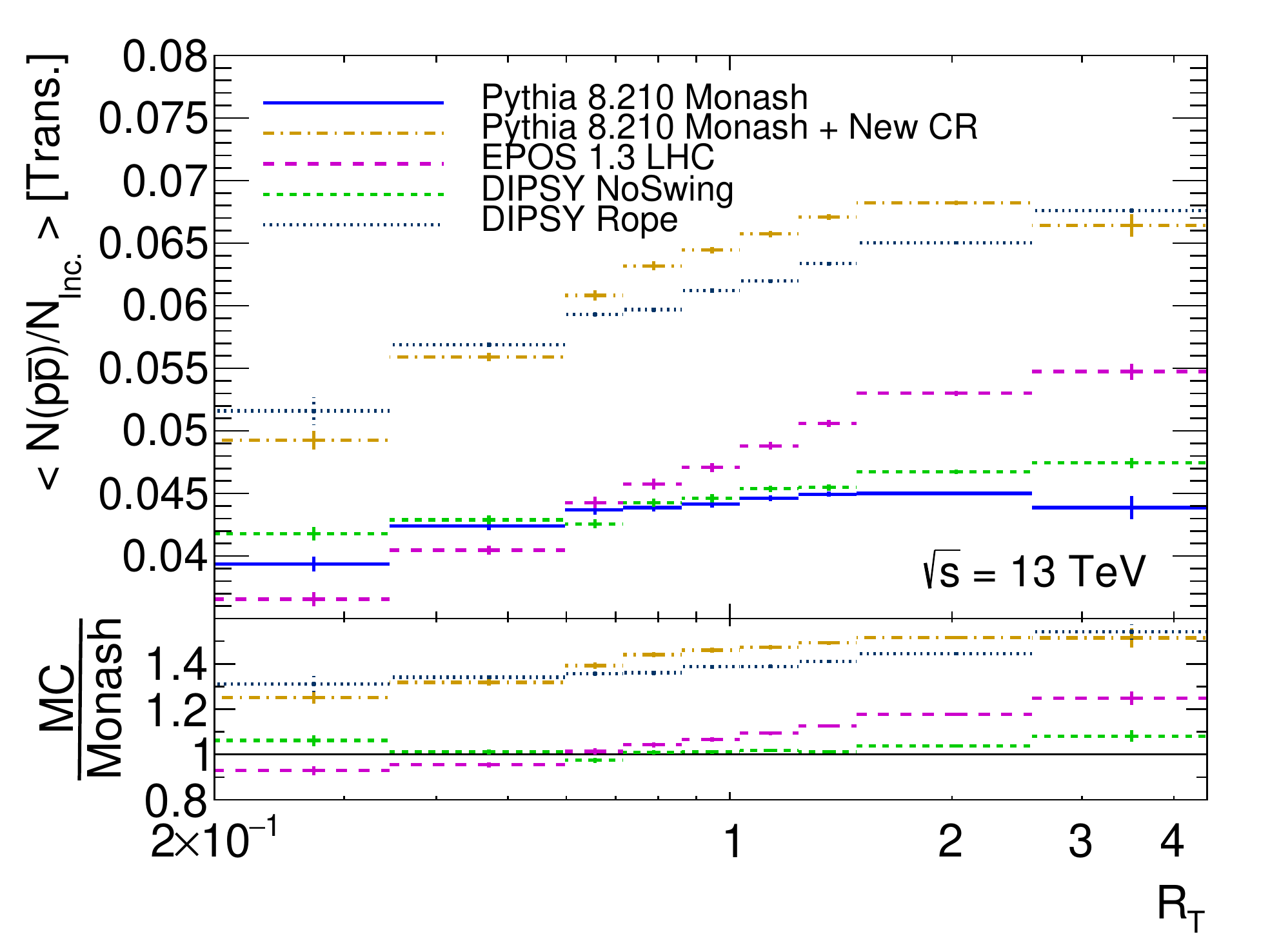}
            \caption{}
            \label{fig:identified-ueB:a}
    \end{subfigure}%
    \begin{subfigure}[b]{0.5\textwidth}
            \centering
            \includegraphics[width=.99\linewidth]{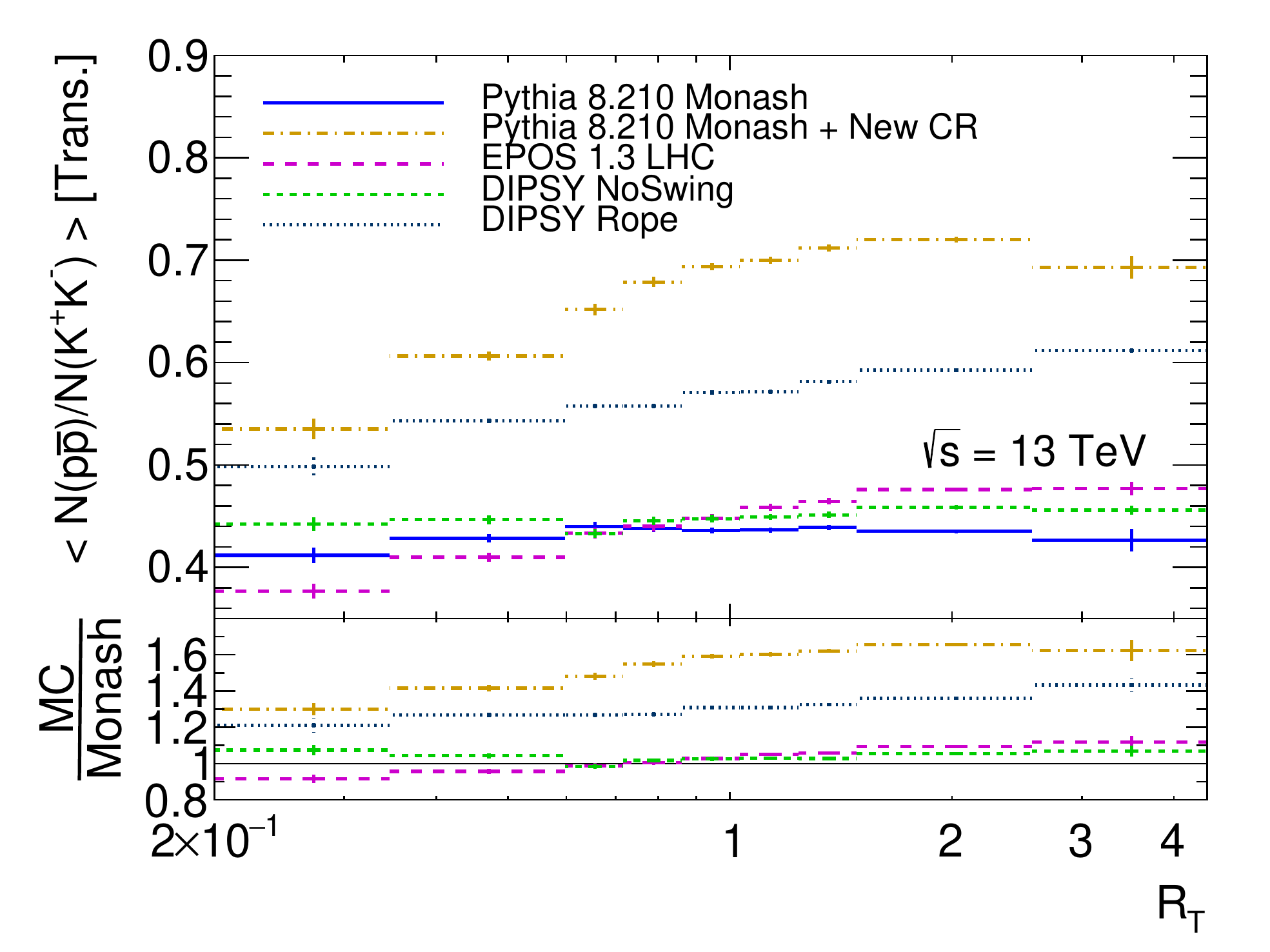}
            \caption{}
            \label{fig:identified-ueB:b}
    \end{subfigure}%
    \newline
    \begin{subfigure}[b]{0.5\textwidth}
            \centering
            \includegraphics[width=.99\linewidth]{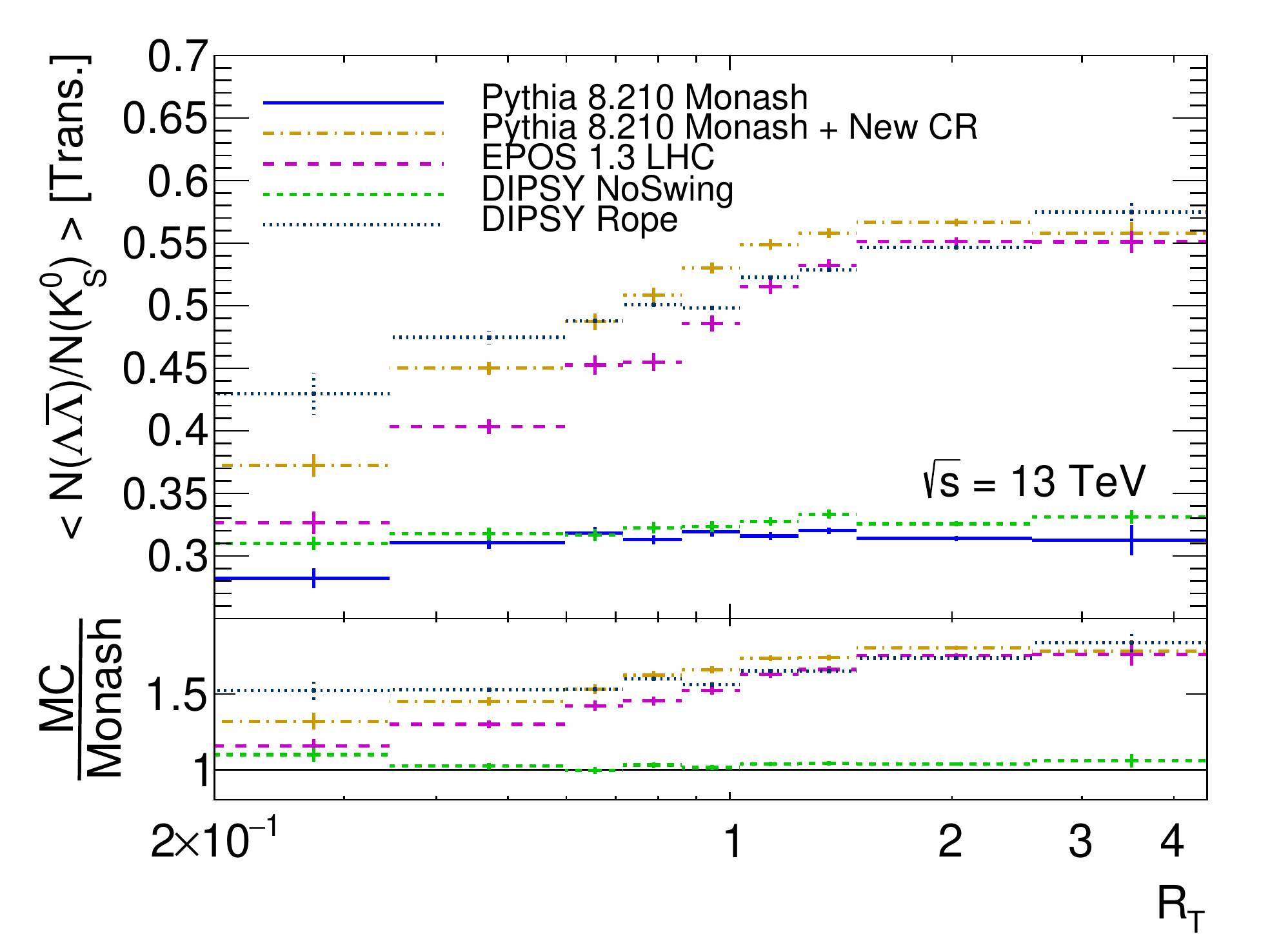}
            \caption{}
            \label{fig:identified-ueB:c}
    \end{subfigure}%
    \begin{subfigure}[b]{0.5\textwidth}
            \centering
            \includegraphics[width=.99\linewidth]{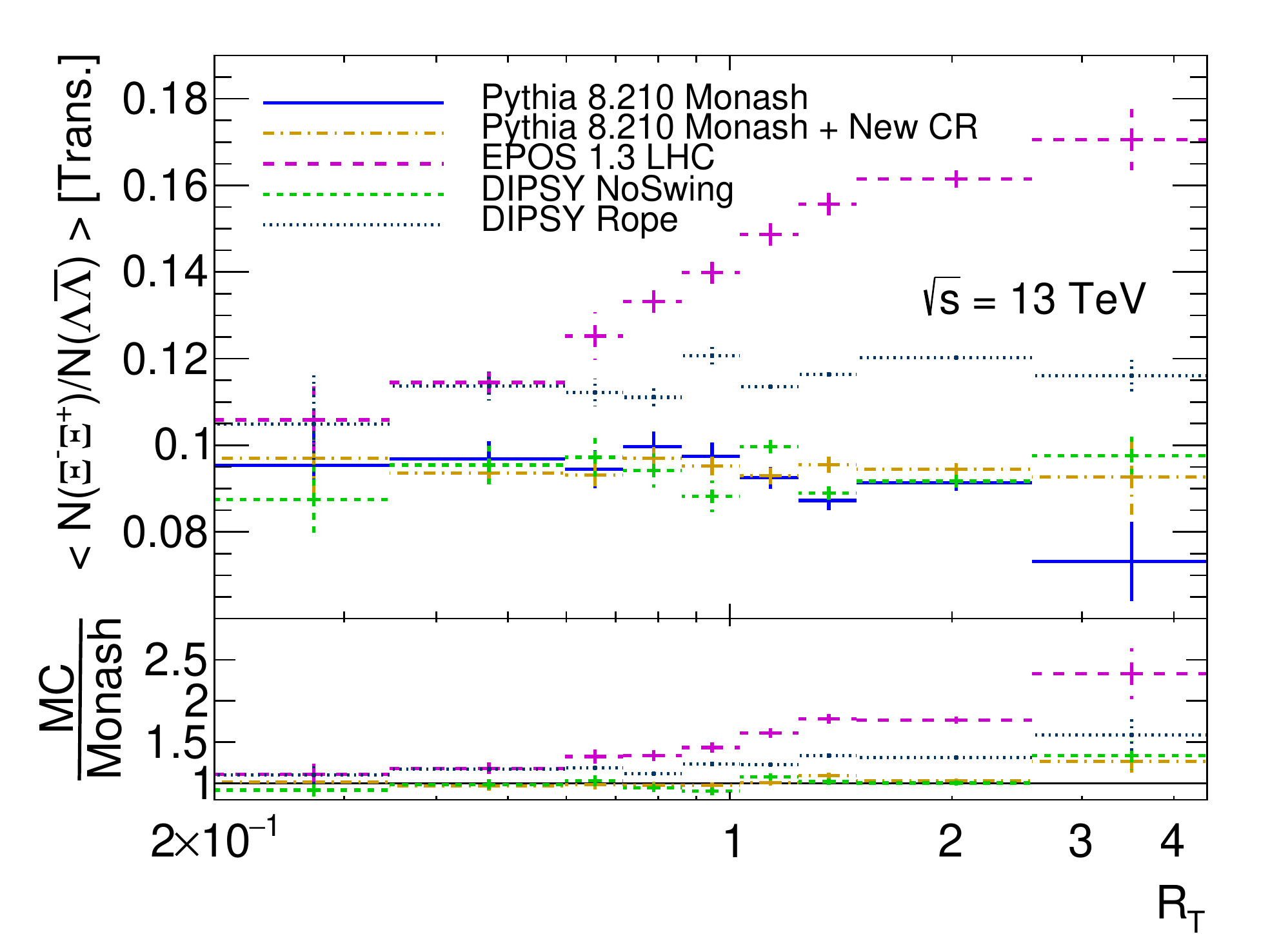}
            \caption{}
            \label{fig:identified-ueB:d}
    \end{subfigure}%
    \caption{Normalised average identified-baryon yields in the transverse region as a function of \Rt for $p\bar{p}/\Nch$ (a), $p\bar{p}/K^+K^-$ (b), $\Lambda\bar{\Lambda}/\kshort$ (c) and $\Xi^+\Xi^-/\Lambda\bar{\Lambda}$ (d).
     Shown for different MC models and tunes. All ratios are relative to \py.210 Monash. Colour online.}
    \label{fig:identified-ueB}
\end{figure}

The baryonic enhancement is probed in \figRef{fig:identified-ueB}. In \figRef{fig:identified-ueB:a} the $p\bar{p}$ multiplicity is normalised to the inclusive charged multiplicity. Some phase-space effects are apparent in all MCs with the smallest rise in proton enhancement coming from \py and \dipsynoswing. In \epos the magnitude of the enhancement is steepest as a function of \Rt leading it to be in agreement with \py and \dipsynoswing only for low values of \Rt. \py + new CR and \dipsyrope both predict significant enhancement, with up to a 50\% increase in proton fraction relative to their simpler model tunes.

By normalising the proton yield to the charged-kaon yield, in \figRef{fig:identified-ueB:b}, we obtain a probe that is sensitive to any relative enhancement between baryons and strangeness. \py Monash + new CR exhibits the highest ratio, as expected since it incorporates a mechanism for baryon enhancement but no mechanism for strangeness enhancement. Next highest is  \dipsyrope, due to a baryon enhancement significantly larger than its strange meson enhancement. We note that in \epos, the two effects are of the same magnitude in both sectors - making it similar to the baseline models, \py and \dipsynoswing in this observable.

In \figRef{fig:identified-ueB:c} strange-baryon enhancement is plotted in the ratio of $\Lambda\bar{\Lambda}$ hyperons to \kshort. The profile is similar to that of the proton enhancement except that all of \epos, \dipsyrope and \py + new CR are in reasonable agreement and show up to a 90\% enhancement effect with increasing \Rt.

Finally doubly-strange $\Xi^\pm$ baryons are investigated in \figRef{fig:identified-ueB:d} normalised to the $\Lambda\bar{\Lambda}$ multiplicity spectra. A clear enhancement with increasing \Rt is observed only for \epos; other models do not predict a strong rise however \dipsyrope predicts a slightly larger baseline fraction of doubly-strange to singly-strange baryons.
\section{Conclusions \label{sec:conclusions}}
A new axis for underlying-event studies $\Rt = \Nch/\left<\Nch\right>$ is proposed to be used as an alternate to leading-object \pt\ for events which contain at least one hard scatter and are hence on the `underlying event plateau'. \Rt\ isolates events with exceptionally large or small activity  in the transverse underlying event region with respect to the event-averaged mean. Distributions which are sensitive to hadronisation effects are studied as a function of this event-activity classifier.

Identified-particle ratios (strangeness and in particular baryon fractions) are observed to be highly sensitive to the event activity density with a marked increase in average $p\bar{p}/\Nch$ and $\Lambda\bar{\Lambda}/\kshort$ multiplicity ratios in dense events for \py with the new colour-reconnection model, \dipsyrope and \epos when compared to models with a less detailed treatment, \py and \dipsynoswing. 

Distributions are also presented for $K$, $\phi$ and $\Xi^\pm$ resonances.

These distributions are experimentally accessible at the LHC and would allow for discrimination between alternate colour-reconnection, colour ``rope'' and hydrodynamic modelling of dense proton-proton interactions. 

\paragraph{Acknowledgements:} We gratefully acknowledge support from the Monash-Warwick Alliance Development Fund, without which this study would not have been possible. 
 PS is the recipient of an Australian Research Council Future Fellowship, FT130100744. This work was also supported in part by the ARC Centre of Excellence for Particle Physics at the Terascale and we acknowledge the support of the UK Science and Technology Facilities Council. We thank C. Bierlich and J. Christiansen for supplying the tunes of \dipsy and \py used in this paper. 

\bibliographystyle{utphys}
\bibliography{main}

\providecommand{\href}[2]{#2}\begingroup\raggedright\begin{thebibliography}{10}

\bibitem{Arnison:1983gw}
{\bfseries UA1} Collaboration, G.~Arnison {\em et~al.}, ``{Hadronic Jet
  Production at the CERN Proton - anti-Proton Collider},''
\href{http://dx.doi.org/10.1016/0370-2693(83)90254-X}{{\em Phys. Lett.}
  {\bfseries B132} (1983) 214}.

\bibitem{Buckley:2011ms}
A.~Buckley {\em et~al.}, ``{General-purpose event generators for LHC
  physics},'' \href{http://dx.doi.org/10.1016/j.physrep.2011.03.005}{{\em Phys.
  Rept.} {\bfseries 504} (2011) 145--233},
\href{http://arxiv.org/abs/1101.2599}{{\ttfamily arXiv:1101.2599 [hep-ph]}}.

\bibitem{Arnison:1982ds}
{\bfseries UA1} Collaboration, G.~Arnison {\em et~al.}, ``{Transverse Energy
  Distributions in the Central Calorimeters},'' in {\em {Proceedings, 21st
  International Conference on High Energy Physics (ICHEP 1982)}}.
\newblock 1982.
\newblock
\url{http://alice.cern.ch/format/showfull?sysnb=0050334}.
\newblock

\bibitem{Arnison:1981ks}
{\bfseries UA1} Collaboration, G.~Arnison {\em et~al.}, ``{Some Observations on
  the First Events Seen at the CERN Proton - anti-Proton Collider},''
  \href{http://dx.doi.org/10.1016/0370-2693(81)90839-X}{{\em Phys. Lett.}
  {\bfseries B107} (1981) 320--324}.
[Erratum: Phys. Lett.109B,510(1982)].

\bibitem{Arnison:1982rm}
{\bfseries UA1} Collaboration, G.~Arnison {\em et~al.}, ``{Charged Particle
  Multiplicity Distributions in Proton Anti-proton Collisions at 540 GeV
  Center-of-mass Energy},''
\href{http://dx.doi.org/10.1016/0370-2693(83)90969-3}{{\em Phys. Lett.}
  {\bfseries B123} (1983) 108}.

\bibitem{Albajar:1988tt}
{\bfseries UA1} Collaboration, C.~Albajar {\em et~al.}, ``{Production of Low
  Transverse Energy Clusters in anti-p p Collisions at $\sqrt{s} = $ 0.2 TeV to
  0.9 TeV and their Interpretation in Terms of QCD Jets},''
\href{http://dx.doi.org/10.1016/0550-3213(88)90450-6}{{\em Nucl. Phys.}
  {\bfseries B309} (1988) 405}.

\bibitem{Caines:2009iy}
{\bfseries STAR} Collaboration, H.~Caines, ``{Exploring Jet Properties in p-p
  Collisions at 200 GeV with STAR},''
  \href{http://dx.doi.org/10.1016/j.nuclphysa.2009.10.096}{{\em Nucl. Phys.}
  {\bfseries A830} (2009) 263C--266C},
\href{http://arxiv.org/abs/0907.3460}{{\ttfamily arXiv:0907.3460 [nucl-ex]}}.

\bibitem{Caines:2011zza}
{\bfseries STAR} Collaboration, H.~Caines, ``{Jet and underlying event
  measurements in p + p collisions at RHIC},''
\href{http://dx.doi.org/10.1016/j.nuclphysa.2011.02.084}{{\em Nucl. Phys.}
  {\bfseries A855} (2011) 376--379}.

\bibitem{Affolder:2001xt}
{\bfseries CDF} Collaboration, T.~Affolder {\em et~al.}, ``{Charged jet
  evolution and the underlying event in $p\bar{p}$ collisions at 1.8 TeV},''
\href{http://dx.doi.org/10.1103/PhysRevD.65.092002}{{\em Phys. Rev.} {\bfseries
  D65} (2002) 092002}.

\bibitem{Acosta:2004wqa}
{\bfseries CDF} Collaboration, D.~Acosta {\em et~al.}, ``{The underlying event
  in hard interactions at the Tevatron $\bar{p}p$ collider},''
  \href{http://dx.doi.org/10.1103/PhysRevD.70.072002}{{\em Phys. Rev.}
  {\bfseries D70} (2004) 072002},
\href{http://arxiv.org/abs/hep-ex/0404004}{{\ttfamily arXiv:hep-ex/0404004
  [hep-ex]}}.

\bibitem{Aaltonen:2010rm}
{\bfseries CDF} Collaboration, T.~Aaltonen {\em et~al.}, ``{Studying the
  Underlying Event in Drell-Yan and High Transverse Momentum Jet Production at
  the Tevatron},'' \href{http://dx.doi.org/10.1103/PhysRevD.82.034001}{{\em
  Phys. Rev.} {\bfseries D82} (2010) 034001},
\href{http://arxiv.org/abs/1003.3146}{{\ttfamily arXiv:1003.3146 [hep-ex]}}.

\bibitem{Aaltonen:2015aoa}
{\bfseries CDF} Collaboration, T.~Aaltonen {\em et~al.}, ``{A Study of the
  Energy Dependence of the Underlying Event in Proton-Antiproton Collisions},''
\href{http://arxiv.org/abs/1508.05340}{{\ttfamily arXiv:1508.05340 [hep-ex]}}.

\bibitem{Khachatryan:2010pv}
{\bfseries CMS} Collaboration, V.~Khachatryan {\em et~al.}, ``{First
  Measurement of the Underlying Event Activity at the LHC with $\sqrt{s} = 0.9$
  TeV},'' \href{http://dx.doi.org/10.1140/epjc/s10052-010-1453-9}{{\em Eur.
  Phys. J.} {\bfseries C70} (2010) 555--572},
\href{http://arxiv.org/abs/1006.2083}{{\ttfamily arXiv:1006.2083 [hep-ex]}}.

\bibitem{Aad:2010fh}
{\bfseries ATLAS} Collaboration, G.~Aad {\em et~al.}, ``{Measurement of
  underlying event characteristics using charged particles in pp collisions at
  $\sqrt{s} = 900$ GeV and 7 TeV with the ATLAS detector},''
  \href{http://dx.doi.org/10.1103/PhysRevD.83.112001}{{\em Phys. Rev.}
  {\bfseries D83} (2011) 112001},
\href{http://arxiv.org/abs/1012.0791}{{\ttfamily arXiv:1012.0791 [hep-ex]}}.

\bibitem{Aad:2011qe}
{\bfseries ATLAS} Collaboration, G.~Aad {\em et~al.}, ``{Measurements of
  underlying-event properties using neutral and charged particles in $pp$
  collisions at 900 GeV and 7 TeV with the ATLAS detector at the LHC},''
  \href{http://dx.doi.org/10.1140/epjc/s10052-011-1636-z}{{\em Eur. Phys. J.}
  {\bfseries C71} (2011) 1636},
\href{http://arxiv.org/abs/1103.1816}{{\ttfamily arXiv:1103.1816 [hep-ex]}}.

\bibitem{Chatrchyan:2011id}
{\bfseries CMS} Collaboration, S.~Chatrchyan {\em et~al.}, ``{Measurement of
  the Underlying Event Activity at the LHC with $\sqrt{s}= 7$ TeV and
  Comparison with $\sqrt{s} = 0.9$ TeV},''
  \href{http://dx.doi.org/10.1007/JHEP09(2011)109}{{\em JHEP} {\bfseries 09}
  (2011) 109},
\href{http://arxiv.org/abs/1107.0330}{{\ttfamily arXiv:1107.0330 [hep-ex]}}.

\bibitem{ALICE:2011ac}
{\bfseries ALICE} Collaboration, B.~Abelev {\em et~al.}, ``{Underlying Event
  measurements in $pp$ collisions at $\sqrt{s}=0.9$ and 7 TeV with the ALICE
  experiment at the LHC},''
  \href{http://dx.doi.org/10.1007/JHEP07(2012)116}{{\em JHEP} {\bfseries 07}
  (2012) 116},
\href{http://arxiv.org/abs/1112.2082}{{\ttfamily arXiv:1112.2082 [hep-ex]}}.

\bibitem{Aad:2012mfa}
{\bfseries ATLAS} Collaboration, G.~Aad {\em et~al.}, ``{Measurements of the
  pseudorapidity dependence of the total transverse energy in proton-proton
  collisions at $\sqrt{s}=7$ TeV with ATLAS},''
  \href{http://dx.doi.org/10.1007/JHEP11(2012)033}{{\em JHEP} {\bfseries 11}
  (2012) 033},
\href{http://arxiv.org/abs/1208.6256}{{\ttfamily arXiv:1208.6256 [hep-ex]}}.

\bibitem{Chatrchyan:2012tt}
{\bfseries CMS} Collaboration, S.~Chatrchyan {\em et~al.}, ``{Measurement of
  the underlying event activity in $pp$ collisions at $\sqrt{s} = 0.9$ and 7
  TeV with the novel jet-area/median approach},''
  \href{http://dx.doi.org/10.1007/JHEP08(2012)130}{{\em JHEP} {\bfseries 08}
  (2012) 130},
\href{http://arxiv.org/abs/1207.2392}{{\ttfamily arXiv:1207.2392 [hep-ex]}}.

\bibitem{Chatrchyan:2012tb}
{\bfseries CMS} Collaboration, S.~Chatrchyan {\em et~al.}, ``{Measurement of
  the underlying event in the Drell-Yan process in proton-proton collisions at
  $\sqrt{s}=7$ TeV},''
  \href{http://dx.doi.org/10.1140/epjc/s10052-012-2080-4}{{\em Eur. Phys. J.}
  {\bfseries C72} (2012) 2080},
\href{http://arxiv.org/abs/1204.1411}{{\ttfamily arXiv:1204.1411 [hep-ex]}}.

\bibitem{Chatrchyan:2013gfi}
{\bfseries CMS} Collaboration, S.~Chatrchyan {\em et~al.}, ``{Study of the
  underlying event at forward rapidity in pp collisions at $\sqrt{s}$ = 0.9,
  2.76, and 7 TeV},'' \href{http://dx.doi.org/10.1007/JHEP04(2013)072}{{\em
  JHEP} {\bfseries 04} (2013) 072},
\href{http://arxiv.org/abs/1302.2394}{{\ttfamily arXiv:1302.2394 [hep-ex]}}.

\bibitem{Chatrchyan:2013ala}
{\bfseries CMS} Collaboration, S.~Chatrchyan {\em et~al.}, ``{Jet and
  underlying event properties as a function of charged-particle multiplicity in
  proton–proton collisions at $\sqrt{s}$ = 7 TeV},''
  \href{http://dx.doi.org/10.1140/epjc/s10052-013-2674-5}{{\em Eur. Phys. J.}
  {\bfseries C73} no.~12, (2013) 2674},
\href{http://arxiv.org/abs/1310.4554}{{\ttfamily arXiv:1310.4554 [hep-ex]}}.

\bibitem{Aad:2014hia}
{\bfseries ATLAS} Collaboration, G.~Aad {\em et~al.}, ``{Measurement of the
  underlying event in jet events from 7 TeV proton-proton collisions with the
  ATLAS detector},''
  \href{http://dx.doi.org/10.1140/epjc/s10052-014-2965-5}{{\em Eur. Phys. J.}
  {\bfseries C74} no.~8, (2014) 2965},
\href{http://arxiv.org/abs/1406.0392}{{\ttfamily arXiv:1406.0392 [hep-ex]}}.

\bibitem{Aad:2014jgf}
{\bfseries ATLAS} Collaboration, G.~Aad {\em et~al.}, ``{Measurement of
  distributions sensitive to the underlying event in inclusive Z-boson
  production in $pp$ collisions at $\sqrt{s}=7$ TeV with the ATLAS detector},''
  \href{http://dx.doi.org/10.1140/epjc/s10052-014-3195-6}{{\em Eur. Phys. J.}
  {\bfseries C74} no.~12, (2014) 3195},
\href{http://arxiv.org/abs/1409.3433}{{\ttfamily arXiv:1409.3433 [hep-ex]}}.

\bibitem{Khachatryan:2015jza}
{\bfseries CMS} Collaboration, V.~Khachatryan {\em et~al.}, ``{Measurement of
  the underlying event activity using charged-particle jets in proton-proton
  collisions at $\sqrt{s} = $ 2.76 TeV},''
  \href{http://dx.doi.org/10.1007/JHEP09(2015)137}{{\em JHEP} {\bfseries 09}
  (2015) 137},
\href{http://arxiv.org/abs/1507.07229}{{\ttfamily arXiv:1507.07229 [hep-ex]}}.

\bibitem{ATL-PHYS-PUB-2015-019}
{\bfseries ATLAS} Collaboration, G.~Aad {\em et~al.}, ``{Leading Track
  Underlying Event at 13 TeV},''. \url{https://cds.cern.ch/record/2037684}.

\bibitem{Cacciari:2009dp}
M.~Cacciari, G.~P. Salam, and S.~Sapeta, ``{On the characterisation of the
  underlying event},'' \href{http://dx.doi.org/10.1007/JHEP04(2010)065}{{\em
  JHEP} {\bfseries 04} (2010) 065},
\href{http://arxiv.org/abs/0912.4926}{{\ttfamily arXiv:0912.4926 [hep-ph]}}.

\bibitem{Heinrich:2011bla}
M.~Heinrich, {\em {A Jet Based Approach to Measuring Soft Contributions to
  Proton-Proton Collisions with the CMS Experiment}}.
\newblock PhD thesis, KIT, Karlsruhe, EKP, 2011.
\newblock
\url{https://inspirehep.net/record/1087971/files/CERN-THESIS-2011-190.pdf}.
\newblock

\bibitem{Bierlich:2015rha}
C.~Bierlich and J.~R. Christiansen, ``{Effects of color reconnection on hadron
  flavor observables},''
  \href{http://dx.doi.org/10.1103/PhysRevD.92.094010}{{\em Phys. Rev.}
  {\bfseries D92} no.~9, (2015) 094010},
\href{http://arxiv.org/abs/1507.02091}{{\ttfamily arXiv:1507.02091 [hep-ph]}}.

\bibitem{Bahr:2008pv}
M.~B{\"a}hr {\em et~al.}, ``{Herwig++ Physics and Manual},''
  \href{http://dx.doi.org/10.1140/epjc/s10052-008-0798-9}{{\em Eur. Phys. J.}
  {\bfseries C58} (2008) 639--707},
\href{http://arxiv.org/abs/0803.0883}{{\ttfamily arXiv:0803.0883 [hep-ph]}}.

\bibitem{Sjostrand:2014zea}
T.~Sj{\"o}strand, S.~Ask, J.~R. Christiansen, R.~Corke, N.~Desai, P.~Ilten,
  S.~Mrenna, S.~Prestel, C.~O. Rasmussen, and P.~Z. Skands, ``{An Introduction
  to PYTHIA 8.2},'' \href{http://dx.doi.org/10.1016/j.cpc.2015.01.024}{{\em
  Comput. Phys. Commun.} {\bfseries 191} (2015) 159--177},
\href{http://arxiv.org/abs/1410.3012}{{\ttfamily arXiv:1410.3012 [hep-ph]}}.

\bibitem{Gleisberg:2008ta}
T.~Gleisberg, S.~Hoeche, F.~Krauss, M.~Schonherr, S.~Schumann, F.~Siegert, and
  J.~Winter, ``{Event generation with SHERPA 1.1},''
  \href{http://dx.doi.org/10.1088/1126-6708/2009/02/007}{{\em JHEP} {\bfseries
  02} (2009) 007},
\href{http://arxiv.org/abs/0811.4622}{{\ttfamily arXiv:0811.4622 [hep-ph]}}.

\bibitem{Karneyeu:2013aha}
A.~Karneyeu, L.~Mijovic, S.~Prestel, and P.~Z. Skands, ``{MCPLOTS: a particle
  physics resource based on volunteer computing},''
  \href{http://dx.doi.org/10.1140/epjc/s10052-014-2714-9}{{\em Eur. Phys. J.}
  {\bfseries C74} (2014) 2714},
\href{http://arxiv.org/abs/1306.3436}{{\ttfamily arXiv:1306.3436 [hep-ph]}}.

\bibitem{Field:2005sa}
{\bfseries CDF} Collaboration, R.~Field and R.~C. Group, ``{PYTHIA tune A,
  HERWIG, and JIMMY in Run 2 at CDF},''
\href{http://arxiv.org/abs/hep-ph/0510198}{{\ttfamily arXiv:hep-ph/0510198
  [hep-ph]}}.

\bibitem{Sjostrand:1987su}
T.~Sj{\"o}strand and M.~van Zijl, ``{A Multiple Interaction Model for the Event
  Structure in Hadron Collisions},''
\href{http://dx.doi.org/10.1103/PhysRevD.36.2019}{{\em Phys. Rev.} {\bfseries
  D36} (1987) 2019}.

\bibitem{Sjostrand:2006za}
T.~Sj{\"o}strand, S.~Mrenna, and P.~Z. Skands, ``{PYTHIA 6.4 Physics and
  Manual},'' \href{http://dx.doi.org/10.1088/1126-6708/2006/05/026}{{\em JHEP}
  {\bfseries 05} (2006) 026},
\href{http://arxiv.org/abs/hep-ph/0603175}{{\ttfamily arXiv:hep-ph/0603175
  [hep-ph]}}.

\bibitem{McLerran:1986nc}
L.~D. McLerran, M.~Kataja, P.~V. Ruuskanen, and H.~von Gersdorff, ``{Studies of
  the Hydrodynamical Evolution of Matter Produced in Fluctuations in p anti-p
  Collisions and in Ultrarelativistic Nuclear Collisions. 2. Transverse
  Momentum Distributions},''
\href{http://dx.doi.org/10.1103/PhysRevD.34.2755}{{\em Phys. Rev.} {\bfseries
  D34} (1986) 2755}.

\bibitem{Albajar:1989an}
{\bfseries UA1} Collaboration, C.~Albajar {\em et~al.}, ``{A Study of the
  General Characteristics of $p\bar{p}$ Collisions at $\sqrt{s}$ = 0.2 TeV to
  0.9 TeV},''
\href{http://dx.doi.org/10.1016/0550-3213(90)90493-W}{{\em Nucl. Phys.}
  {\bfseries B335} (1990) 261--287}.

\bibitem{Acosta:2001rm}
{\bfseries CDF} Collaboration, D.~Acosta {\em et~al.}, ``{Soft and hard
  interactions in $p\bar{p}$ collisions at $\sqrt{s}=$ 1800 GeV and 630 GeV},''
\href{http://dx.doi.org/10.1103/PhysRevD.65.072005}{{\em Phys. Rev.} {\bfseries
  D65} (2002) 072005}.

\bibitem{Aaltonen:2009ne}
{\bfseries CDF} Collaboration, T.~Aaltonen {\em et~al.}, ``{Measurement of
  Particle Production and Inclusive Differential Cross Sections in $p \bar{p}$
  Collisions at $\sqrt{s} = 1.96$-TeV},''
  \href{http://dx.doi.org/10.1103/PhysRevD.82.119903,
  10.1103/PhysRevD.79.112005}{{\em Phys. Rev.} {\bfseries D79} (2009) 112005},
  \href{http://arxiv.org/abs/0904.1098}{{\ttfamily arXiv:0904.1098 [hep-ex]}}.
[Erratum: Phys. Rev.D82,119903(2010)].

\bibitem{Aad:2010ac}
{\bfseries ATLAS} Collaboration, G.~Aad {\em et~al.}, ``{Charged-particle
  multiplicities in pp interactions measured with the ATLAS detector at the
  LHC},'' \href{http://dx.doi.org/10.1088/1367-2630/13/5/053033}{{\em New J.
  Phys.} {\bfseries 13} (2011) 053033},
\href{http://arxiv.org/abs/1012.5104}{{\ttfamily arXiv:1012.5104 [hep-ex]}}.

\bibitem{Khachatryan:2010nk}
{\bfseries CMS} Collaboration, V.~Khachatryan {\em et~al.}, ``{Charged particle
  multiplicities in $pp$ interactions at $\sqrt{s}=0.9$, 2.36, and 7 TeV},''
  \href{http://dx.doi.org/10.1007/JHEP01(2011)079}{{\em JHEP} {\bfseries 01}
  (2011) 079},
\href{http://arxiv.org/abs/1011.5531}{{\ttfamily arXiv:1011.5531 [hep-ex]}}.

\bibitem{Adams:2006yu}
{\bfseries STAR} Collaboration, B.~I. Abelev {\em et~al.}, ``{Strange baryon
  resonance production in $\sqrt{s}$(NN) = 200 GeV p+p and Au+Au collisions},''
  \href{http://dx.doi.org/10.1103/PhysRevLett.97.132301}{{\em Phys. Rev. Lett.}
  {\bfseries 97} (2006) 132301},
\href{http://arxiv.org/abs/nucl-ex/0604019}{{\ttfamily arXiv:nucl-ex/0604019
  [nucl-ex]}}.

\bibitem{Aaltonen:2011wz}
{\bfseries CDF} Collaboration, T.~Aaltonen {\em et~al.}, ``{Production of
  $\Lambda$, $\bar{\Lambda}^0$ $\Xi^{\pm}$ and $\Omega^{\pm}$ Hyperons in $p
  \bar{p}$ Collisions at $\sqrt{s} = 1.96$ TeV},''
  \href{http://dx.doi.org/10.1103/PhysRevD.86.012002}{{\em Phys. Rev.}
  {\bfseries D86} (2012) 012002},
\href{http://arxiv.org/abs/1101.2996}{{\ttfamily arXiv:1101.2996 [hep-ex]}}.

\bibitem{Aad:2011hd}
{\bfseries ATLAS} Collaboration, G.~Aad {\em et~al.}, ``{Kshort and $\Lambda$
  production in $pp$ interactions at $\sqrt{s}=0.9$ and 7 TeV measured with the
  ATLAS detector at the LHC},''
  \href{http://dx.doi.org/10.1103/PhysRevD.85.012001}{{\em Phys. Rev.}
  {\bfseries D85} (2012) 012001},
\href{http://arxiv.org/abs/1111.1297}{{\ttfamily arXiv:1111.1297 [hep-ex]}}.

\bibitem{Khachatryan:2011tm}
{\bfseries CMS} Collaboration, V.~Khachatryan {\em et~al.}, ``{Strange Particle
  Production in $pp$ Collisions at $\sqrt{s}=0.9$ and 7 TeV},''
  \href{http://dx.doi.org/10.1007/JHEP05(2011)064}{{\em JHEP} {\bfseries 05}
  (2011) 064},
\href{http://arxiv.org/abs/1102.4282}{{\ttfamily arXiv:1102.4282 [hep-ex]}}.

\bibitem{Aaij:2012ut}
{\bfseries LHCb} Collaboration, R.~Aaij {\em et~al.}, ``{Measurement of prompt
  hadron production ratios in $pp$ collisions at $\sqrt{s} = $ 0.9 and 7
  TeV},'' \href{http://dx.doi.org/10.1140/epjc/s10052-012-2168-x}{{\em Eur.
  Phys. J.} {\bfseries C72} (2012) 2168},
\href{http://arxiv.org/abs/1206.5160}{{\ttfamily arXiv:1206.5160 [hep-ex]}}.

\bibitem{Abelev:2012jp}
{\bfseries ALICE} Collaboration, B.~Abelev {\em et~al.}, ``{Multi-strange
  baryon production in $pp$ collisions at $\sqrt{s} = 7$ TeV with ALICE},''
  \href{http://dx.doi.org/10.1016/j.physletb.2012.05.011}{{\em Phys. Lett.}
  {\bfseries B712} (2012) 309--318},
\href{http://arxiv.org/abs/1204.0282}{{\ttfamily arXiv:1204.0282 [nucl-ex]}}.

\bibitem{Chatrchyan:2013qsa}
{\bfseries CMS} Collaboration, S.~Chatrchyan {\em et~al.}, ``{Measurement of
  neutral strange particle production in the underlying event in proton-proton
  collisions at $\sqrt{s} = $7 TeV},''
  \href{http://dx.doi.org/10.1103/PhysRevD.88.052001}{{\em Phys. Rev.}
  {\bfseries D88} (2013) 052001},
\href{http://arxiv.org/abs/1305.6016}{{\ttfamily arXiv:1305.6016 [hep-ex]}}.

\bibitem{Abelev:2014qqa}
{\bfseries ALICE} Collaboration, B.~B. Abelev {\em et~al.}, ``{Production of
  $\Sigma(1385)^{\pm}$ and $\Xi(1530)^{0}$ in proton-proton collisions at
  $\sqrt{s}=$ 7 TeV},''
  \href{http://dx.doi.org/10.1140/epjc/s10052-014-3191-x}{{\em Eur. Phys. J.}
  {\bfseries C75} no.~1, (2015) 1},
\href{http://arxiv.org/abs/1406.3206}{{\ttfamily arXiv:1406.3206 [nucl-ex]}}.

\bibitem{Abelev:2006cs}
{\bfseries STAR} Collaboration, B.~I. Abelev {\em et~al.}, ``{Strange particle
  production in p+p collisions at $\sqrt{s} =$ 200 GeV},''
  \href{http://dx.doi.org/10.1103/PhysRevC.75.064901}{{\em Phys. Rev.}
  {\bfseries C75} (2007) 064901},
\href{http://arxiv.org/abs/nucl-ex/0607033}{{\ttfamily arXiv:nucl-ex/0607033
  [nucl-ex]}}.

\bibitem{Aamodt:2011zj}
{\bfseries ALICE} Collaboration, K.~Aamodt {\em et~al.}, ``{Production of
  pions, kaons and protons in $pp$ collisions at $\sqrt{s}= 900$ GeV with ALICE
  at the LHC},'' \href{http://dx.doi.org/10.1140/epjc/s10052-011-1655-9}{{\em
  Eur. Phys. J.} {\bfseries C71} (2011) 1655},
\href{http://arxiv.org/abs/1101.4110}{{\ttfamily arXiv:1101.4110 [hep-ex]}}.

\bibitem{Abelev:2012hy}
{\bfseries ALICE} Collaboration, B.~Abelev {\em et~al.}, ``{Production of
  $K^*(892)^0$ and $\phi(1020)$ in $pp$ collisions at $\sqrt{s}=7$ TeV},''
  \href{http://dx.doi.org/10.1140/epjc/s10052-012-2183-y}{{\em Eur. Phys. J.}
  {\bfseries C72} (2012) 2183},
\href{http://arxiv.org/abs/1208.5717}{{\ttfamily arXiv:1208.5717 [hep-ex]}}.

\bibitem{Adam:2015qaa}
{\bfseries ALICE} Collaboration, J.~Adam {\em et~al.}, ``{Measurement of pion,
  kaon and proton production in proton–proton collisions at $\sqrt{s} = 7$
  TeV},'' \href{http://dx.doi.org/10.1140/epjc/s10052-015-3422-9}{{\em Eur.
  Phys. J.} {\bfseries C75} no.~5, (2015) 226},
\href{http://arxiv.org/abs/1504.00024}{{\ttfamily arXiv:1504.00024 [nucl-ex]}}.

\bibitem{Khachatryan:2010gv}
{\bfseries CMS} Collaboration, V.~Khachatryan {\em et~al.}, ``{Observation of
  Long-Range Near-Side Angular Correlations in Proton-Proton Collisions at the
  LHC},'' \href{http://dx.doi.org/10.1007/JHEP09(2010)091}{{\em JHEP}
  {\bfseries 09} (2010) 091},
\href{http://arxiv.org/abs/1009.4122}{{\ttfamily arXiv:1009.4122 [hep-ex]}}.

\bibitem{Aad:2015gqa}
{\bfseries ATLAS} Collaboration, G.~Aad {\em et~al.}, ``{Observation of
  long-range elliptic anisotropies in $\sqrt{s}=$13 and 2.76 TeV $pp$
  collisions with the ATLAS detector},''
\href{http://arxiv.org/abs/1509.04776}{{\ttfamily arXiv:1509.04776 [hep-ex]}}.

\bibitem{Acosta:2005pk}
{\bfseries CDF} Collaboration, D.~Acosta {\em et~al.}, ``{$K^0_S$ and
  $\Lambda^0$ production studies in $p\bar{p}$ collisions at $\sqrt{s}=$ 1800
  GeV and 630 GeV},'' \href{http://dx.doi.org/10.1103/PhysRevD.72.052001}{{\em
  Phys. Rev.} {\bfseries D72} (2005) 052001},
\href{http://arxiv.org/abs/hep-ex/0504048}{{\ttfamily arXiv:hep-ex/0504048
  [hep-ex]}}.

\bibitem{Abelev:2013sqa}
{\bfseries ALICE} Collaboration, B.~Abelev {\em et~al.}, ``{Multiplicity
  dependence of two-particle azimuthal correlations in pp collisions at the
  LHC},'' \href{http://dx.doi.org/10.1007/JHEP09(2013)049}{{\em JHEP}
  {\bfseries 09} (2013) 049},
\href{http://arxiv.org/abs/1307.1249}{{\ttfamily arXiv:1307.1249 [nucl-ex]}}.

\bibitem{Ortiz:2016mra}
A.~Ortiz, G.~Bencédi, H.~Bello, and S.~Jena, ``{Jet effects in
  high-multiplicity pp events},'' in {\em {7th International Workshop on
  Multiple Partonic Interactions at the LHC (MPI@LHC 2015) Miramare, Trieste,
  Italy, November 23-27, 2015}}.
\newblock 2016.
\newblock
\href{http://arxiv.org/abs/1603.05213}{{\ttfamily arXiv:1603.05213 [hep-ph]}}.
\newblock

\bibitem{Skands:2014pea}
P.~Skands, S.~Carrazza, and J.~Rojo, ``{Tuning PYTHIA 8.1: the Monash 2013
  Tune},'' \href{http://dx.doi.org/10.1140/epjc/s10052-014-3024-y}{{\em Eur.
  Phys. J.} {\bfseries C74} no.~8, (2014) 3024},
\href{http://arxiv.org/abs/1404.5630}{{\ttfamily arXiv:1404.5630 [hep-ph]}}.

\bibitem{Buckley:2010ar}
A.~Buckley, J.~Butterworth, L.~Lonnblad, D.~Grellscheid, H.~Hoeth, J.~Monk,
  H.~Schulz, and F.~Siegert, ``{Rivet user manual},''
  \href{http://dx.doi.org/10.1016/j.cpc.2013.05.021}{{\em Comput. Phys.
  Commun.} {\bfseries 184} (2013) 2803--2819},
\href{http://arxiv.org/abs/1003.0694}{{\ttfamily arXiv:1003.0694 [hep-ph]}}.

\bibitem{Cacciari:2011ma}
M.~Cacciari, G.~P. Salam, and G.~Soyez, ``{FastJet User Manual},''
  \href{http://dx.doi.org/10.1140/epjc/s10052-012-1896-2}{{\em Eur. Phys. J.}
  {\bfseries C72} (2012) 1896},
\href{http://arxiv.org/abs/1111.6097}{{\ttfamily arXiv:1111.6097 [hep-ph]}}.

\bibitem{Sjostrand:2004ef}
T.~Sj{\"o}strand and P.~Z. Skands, ``{Transverse-momentum-ordered showers and
  interleaved multiple interactions},''
  \href{http://dx.doi.org/10.1140/epjc/s2004-02084-y}{{\em Eur. Phys. J.}
  {\bfseries C39} (2005) 129--154},
\href{http://arxiv.org/abs/hep-ph/0408302}{{\ttfamily arXiv:hep-ph/0408302
  [hep-ph]}}.

\bibitem{Sjostrand:2004pf}
T.~Sj{\"o}strand and P.~Z. Skands, ``{Multiple interactions and the structure
  of beam remnants},''
  \href{http://dx.doi.org/10.1088/1126-6708/2004/03/053}{{\em JHEP} {\bfseries
  03} (2004) 053},
\href{http://arxiv.org/abs/hep-ph/0402078}{{\ttfamily arXiv:hep-ph/0402078
  [hep-ph]}}.

\bibitem{Argyropoulos:2014zoa}
S.~Argyropoulos and T.~Sjöstrand, ``{Effects of color reconnection on
  $t\bar{t}$ final states at the LHC},''
  \href{http://dx.doi.org/10.1007/JHEP11(2014)043}{{\em JHEP} {\bfseries 11}
  (2014) 043},
\href{http://arxiv.org/abs/1407.6653}{{\ttfamily arXiv:1407.6653 [hep-ph]}}.

\bibitem{Christiansen:2015yqa}
J.~R. Christiansen and P.~Z. Skands, ``{String Formation Beyond Leading
  Colour},'' \href{http://dx.doi.org/10.1007/JHEP08(2015)003}{{\em JHEP}
  {\bfseries 08} (2015) 003},
\href{http://arxiv.org/abs/1505.01681}{{\ttfamily arXiv:1505.01681 [hep-ph]}}.

\bibitem{Corke:2010yf}
R.~Corke and T.~Sj{\"o}strand, ``{Interleaved Parton Showers and Tuning
  Prospects},'' \href{http://dx.doi.org/10.1007/JHEP03(2011)032}{{\em JHEP}
  {\bfseries 03} (2011) 032},
\href{http://arxiv.org/abs/1011.1759}{{\ttfamily arXiv:1011.1759 [hep-ph]}}.

\bibitem{Sjostrand:2002ip}
T.~Sj{\"o}strand and P.~Z. Skands, ``{Baryon number violation and string
  topologies},'' \href{http://dx.doi.org/10.1016/S0550-3213(03)00193-7}{{\em
  Nucl. Phys.} {\bfseries B659} (2003) 243},
\href{http://arxiv.org/abs/hep-ph/0212264}{{\ttfamily arXiv:hep-ph/0212264
  [hep-ph]}}.

\bibitem{Flensburg:2011kk}
C.~Flensburg, G.~Gustafson, and L.~L{\"o}nnblad, ``{Inclusive and Exclusive
  Observables from Dipoles in High Energy Collisions},''
  \href{http://dx.doi.org/10.1007/JHEP08(2011)103}{{\em JHEP} {\bfseries 08}
  (2011) 103},
\href{http://arxiv.org/abs/1103.4321}{{\ttfamily arXiv:1103.4321 [hep-ph]}}.

\bibitem{Mueller:1993rr}
A.~H. Mueller, ``{Soft gluons in the infinite momentum wave function and the
  BFKL pomeron},''
\href{http://dx.doi.org/10.1016/0550-3213(94)90116-3}{{\em Nucl. Phys.}
  {\bfseries B415} (1994) 373--385}.

\bibitem{Bierlich:2014xba}
C.~Bierlich, G.~Gustafson, L.~Lönnblad, and A.~Tarasov, ``{Effects of
  Overlapping Strings in pp Collisions},''
  \href{http://dx.doi.org/10.1007/JHEP03(2015)148}{{\em JHEP} {\bfseries 03}
  (2015) 148},
\href{http://arxiv.org/abs/1412.6259}{{\ttfamily arXiv:1412.6259 [hep-ph]}}.

\bibitem{Werner:2005jf}
K.~Werner, F.-M. Liu, and T.~Pierog, ``{Parton ladder splitting and the
  rapidity dependence of transverse momentum spectra in deuteron-gold
  collisions at RHIC},''
  \href{http://dx.doi.org/10.1103/PhysRevC.74.044902}{{\em Phys. Rev.}
  {\bfseries C74} (2006) 044902},
\href{http://arxiv.org/abs/hep-ph/0506232}{{\ttfamily arXiv:hep-ph/0506232
  [hep-ph]}}.

\bibitem{Drescher:2000ha}
H.~J. Drescher, M.~Hladik, S.~Ostapchenko, T.~Pierog, and K.~Werner, ``{Parton
  based Gribov-Regge theory},''
  \href{http://dx.doi.org/10.1016/S0370-1573(00)00122-8}{{\em Phys. Rept.}
  {\bfseries 350} (2001) 93--289},
\href{http://arxiv.org/abs/hep-ph/0007198}{{\ttfamily arXiv:hep-ph/0007198
  [hep-ph]}}.

\bibitem{Werner:2007bf}
K.~Werner, ``{Core-corona separation in ultra-relativistic heavy ion
  collisions},'' \href{http://dx.doi.org/10.1103/PhysRevLett.98.152301}{{\em
  Phys. Rev. Lett.} {\bfseries 98} (2007) 152301},
\href{http://arxiv.org/abs/0704.1270}{{\ttfamily arXiv:0704.1270 [nucl-th]}}.

\bibitem{Pierog:2013ria}
T.~Pierog, I.~Karpenko, J.~M. Katzy, E.~Yatsenko, and K.~Werner, ``{EPOS LHC:
  Test of collective hadronization with data measured at the CERN Large Hadron
  Collider},'' \href{http://dx.doi.org/10.1103/PhysRevC.92.034906}{{\em Phys.
  Rev.} {\bfseries C92} no.~3, (2015) 034906},
\href{http://arxiv.org/abs/1306.0121}{{\ttfamily arXiv:1306.0121 [hep-ph]}}.

\bibitem{Aad:2016mok}
{\bfseries ATLAS} Collaboration, G.~Aad {\em et~al.}, ``{Charged-particle
  distributions in $\sqrt{s}=13$ TeV $pp$ interactions measured with the ATLAS
  detector at the LHC},''
\href{http://arxiv.org/abs/1602.01633}{{\ttfamily arXiv:1602.01633 [hep-ex]}}.

\bibitem{Cacciari:2008gp}
M.~Cacciari, G.~P. Salam, and G.~Soyez, ``{The Anti-k(t) jet clustering
  algorithm},'' \href{http://dx.doi.org/10.1088/1126-6708/2008/04/063}{{\em
  JHEP} {\bfseries 04} (2008) 063},
\href{http://arxiv.org/abs/0802.1189}{{\ttfamily arXiv:0802.1189 [hep-ph]}}.

\bibitem{Koba:1972ng}
Z.~Koba, H.~B. Nielsen, and P.~Olesen, ``{Scaling of multiplicity distributions
  in high-energy hadron collisions},''
\href{http://dx.doi.org/10.1016/0550-3213(72)90551-2}{{\em Nucl. Phys.}
  {\bfseries B40} (1972) 317--334}.

\bibitem{CMS:2012qk}
{\bfseries CMS} Collaboration, S.~Chatrchyan {\em et~al.}, ``{Observation of
  long-range near-side angular correlations in proton-lead collisions at the
  LHC},'' \href{http://dx.doi.org/10.1016/j.physletb.2012.11.025}{{\em Phys.
  Lett.} {\bfseries B718} (2013) 795--814},
\href{http://arxiv.org/abs/1210.5482}{{\ttfamily arXiv:1210.5482 [nucl-ex]}}.

\end{thebibliography}\endgroup

\end{document}